\documentclass[pra,reprint,twocolumn,superscriptaddress,showpacs,floatfix]{revtex4-1}
\usepackage{amsmath}
\usepackage{mathrsfs}
\usepackage{txfonts}
\usepackage{amssymb}
\usepackage{graphicx}
\usepackage{hyperref}
\usepackage{ulem}
 \usepackage{overpic}
 \usepackage{psfrag}
\usepackage{tabularx}
\usepackage{multirow}
\usepackage{array}
\usepackage{placeins}
\newcommand{\PreserveBackslash}[1]{\let\temp=\\#1\let\\=\temp}

\renewcommand{\v}[1]{{\bf #1}}
\newcommand{\nn}{\nonumber \\}

\makeatletter

\begin{document}

\title{Entanglement and logarithmic spirals in a quantum spin-1 many-body system with competing dimer and trimer interactions}

\author{Huan-Qiang Zhou}
\affiliation{Centre for Modern Physics, Chongqing University, Chongqing 400044, The People's Republic of China}

\author{Qian-Qian Shi}
\affiliation{Centre for Modern Physics, Chongqing University, Chongqing 400044, The People's Republic of China}

\author{Ian P. McCulloch}
\affiliation{ Department of Physics, National Tsing Hua University, Hsinchu 30013, Taiwan}
\affiliation{Centre for Modern Physics, Chongqing University, Chongqing 400044, The People's Republic of China}

\author{Murray T. Batchelor}
\affiliation{Mathematical Sciences Institute, The Australian National University, Canberra ACT 2601, Australia}
\affiliation{Centre for Modern Physics, Chongqing University, Chongqing 400044, The People's Republic of China}

\begin{abstract}
Spontaneous symmetry breaking (SSB) with type-B Goldstone modes is investigated in the macroscopically degenerate phase  for a quantum spin-1 many-body system with competing dimer and trimer interactions. The SSB involves three distinct patterns. The first occurs at the dimer point, with the pattern from staggered ${\rm SU}(3)$  to ${\rm U}(1)\times{\rm U}(1)$. The second  occurs at the trimer point, with the pattern from uniform ${\rm SU}(3)$ to ${\rm U}(1)\times{\rm U}(1)$. The third occurs in the dimer-trimer regime, with the pattern from uniform ${\rm SU}(2)$ to ${\rm U}(1)$. The number of type-B Goldstone modes is thus two, two and one for the three patterns, respectively. The ground state degeneracies arising from the three patterns are exponential with the system size, which may be recognized as sequences of integers relevant to self-similar logarithmic spirals. This in turn is attributed to the presence of an emergent symmetry operation tailored to a specific degenerate ground state. As a consequence, the residual entropy is non-zero, which measures the disorder present in a unit cell of highly degenerate ground state generated from a generalized highest weight state.
An exact Schmidt decomposition exists for the highly degenerate ground states, thus exposing the self-similarities underlying an abstract fractal, described by the fractal dimension. The latter is extracted from  performing a universal finite system-size scaling analysis of the entanglement entropy, which is identical to the number of type-B Goldstone modes. The model under investigation thus accommodates an exotic scale invariant quantum state of matter.
\end{abstract}

\maketitle
\section{Introduction}
Over many years, significant progress has been made towards  understanding the fundamental notion of spontaneous symmetry breaking (SSB),  with a focus on the classification of the Goldstone modes (GMs) ~\cite{goldstone,Nambu1,Hnielsen, schafer, miransky, nambu, nicolis, brauner1, brauner-watanabe, watanabe, NG,NG1}. The introduction of type-A and type-B GMs~\cite{watanabe} partially resolved the debate between Anderson and Peierls regarding whether or not ${\rm SU}(2)$ ferromagnetic states result from SSB~\cite{anderson, peierls}.
However, not many  quantum many-body systems in condensed matter physics are known to exhibit SSB with type-B GMs, with {only} a few exceptional recent examples~\cite{FMGM,LLspin1,golden,SU4}, in addition to the paradigmatic example of the ${\rm SU}(2)$ spin-$s$ ferromagnetic Heisenberg model. This is in sharp contrast to SSB with type-A GMs which occupies a prominent position in the interpretation of many fundamental phenomena in diverse areas of physics. In fact, a systematic investigation into a variety of quantum many-body systems undergoing SSB with type-A GMs has been underway for the past decade~\cite{Kallin,Song, Metlitski,squarecubic,bilayer,typeAGM, alet, bauer}.
It is therefore also highly desirable to search for quantum many-body systems exhibiting SSB with type-B GMs.

As we have learned from the known examples, quantum many-body systems exhibiting SSB with type-B GMs are exactly solvable, as far as the ground state subspace is concerned. This is due to the observation that all known models undergoing SSB with ${\it only}$ type-B GMs are frustration-free~\cite{tasaki}. Indeed, some of them are even completely integrable in the sense of the quantum Yang-Baxter equation~\cite{baxterbook,sutherlandb,mccoy}, or the Templerley-Lieb algebra~\cite{tla,baxterbook,martin}, if one restricts to their one-dimensional versions. In addition, there is an unexpected connection with the frustration-free models via Witten's conjugation~\cite{katsura}. The models yield highly degenerate ground states, which are scale invariant but not conformally invariant~\cite{FMGM,LLspin1,golden,SU4}. Meanwhile, highly degenerate ground states are subject to an exact Schmidt decomposition~\cite{svd}. It follows that self-similarities underly the ground state subspace, which reflect an abstract fractal~\cite{FMGM,LLspin1,golden,SU4,finitesize}. Indeed, such an abstract fractal is characterized in terms of the fractal dimension, first introduced by  Castro-Alvaredo and Doyon for the  ${\rm SU}(2)$ ferromagnetic states~\cite{doyon} (see also Ref.~\cite{popkov}). Meanwhile, a systematic finite block-size scaling analysis of the entanglement entropy in the thermodynamic limit leads to the identification of the number of type-B GMs with the fractal dimension~\cite{FMGM}.

Given that highly degenerate ground states may be regarded as a characteristic feature for SSB with type-B GMs, it is plausible to search for  quantum many-body systems with the ground state degeneracies being polynomial or exponential with the system size, to scrutinize them as possible candidates for SSB with type-B GMs. This is particularly so  if the ground state degeneracy depends on what types of boundary conditions are adopted, i.e., open boundary conditions (OBCs) and  periodic boundary conditions (PBCs), when the degeneracy is exponential as a function of the system size. 
With this observation as a guiding principle, we choose as an illustrative example, a quantum spin-1 many-body system with competing dimer and trimer interactions, introduced in Ref.~\cite{dtmodel}.

In this paper, we focus on the macroscopically degenerate (MD) phase between the two quantum phase transition (QPT) points in the above-mentioned spin-1 dimer-trimer model~\cite{dtmodel,dtmodel2}.
Indeed, the ground state degeneracies are exponential  as a function of the system size. This is also true at each endpoint of the MD phase, namely at the dimer point and at the trimer point, which are regarded as QPT  points. As a convention, we refer to the MD regime to include the two QPT points, and refer to the dimer-trimer regime to exclude them.
As it turns out, the model exhibits three distinct SSB patterns. The first occurs at the dimer point, with the SSB pattern from the staggered symmetry group ${\rm SU}(3)$  to ${\rm U}(1)\times{\rm U}(1)$, so the number of type-B GMs is two. The second  occurs at the trimer point, with the SSB pattern from the  uniform symmetry group ${\rm SU}(3)$ to ${\rm U}(1)\times{\rm U}(1)$, so the number of type-B GMs is two. The third occurs in the dimer-trimer regime, with the SSB pattern from the uniform symmetry group ${\rm SU}(2)$ to ${\rm U}(1)$, with the number of type-B GMs being one. The ground state degeneracies arising from the three SSB patterns are exponential with the system size, which may be recognized as sequences of integers relevant to self-similar geometric objects - logarithmic spirals.  As a consequence, the residual entropy $S_{\!r}$ is non-zero.

Meanwhile, highly degenerate ground states as a result of the three SSB patterns admit an exact Schmidt decomposition, thus exposing the self-similarities. The latter in turn lead to an abstract fractal underlying the ground state subspace. The fractal dimension for such an abstract fractal is extracted by performing a universal finite system-size scaling analysis of the entanglement entropy. In fact, the finite-size scaling relation reduces to a logarithmic scaling relation with the block size in the thermodynamic limit, with the prefactor being half the fractal dimension, which turns out to be identical to the number of type-B GMs.

 As a result of the three SSB patterns, the spin-1 dimer-trimer model accommodates  scale invariant, but not conformally invariant, ground states that represent an exotic quantum state of matter, featuring that the translation-invariant ground states coexist with $p$-merized ground states for integer $p$. A $p$-merized state is a ground state invariant under the $p$-site translation operation. In the dimer-trimer regime and at the trimer point, $p \ge 2$.  
 In contrast, at the dimer point, $p$ is even~\cite{golden}.

 The paper is organized as follows. In Section~\ref{mh}, we introduce the model Hamiltonian and discuss the ground state phase diagram for the spin-1 dimer-trimer model~\cite{dtmodel}. In Section~\ref{gss}, the ground state subspaces and the ground state degeneracies are discussed in the MD regime, including the two endpoints. In particular, an emergent (local) symmetry operation tailored to a specific ground state is revealed for the three SSB patterns, which is  responsible for the fact that the ground state degeneracies are exponential with the system size. As a result the residual entropy is non-zero, measuring the disorder present in degenerate ground states arising from generalized highest weight states. In Section~\ref{svd}, an exact Schmidt decomposition reflecting self-similarities underlying an abstract fractal in the ground state subspace is exposed for the highly degenerate ground states. In Section~\ref{ee}, a universal finite system-size scaling analysis is performed for the highly degenerate ground states arising from the three SSB patterns, with a focus on the dimer-trimer regime and the trimer point. As a result, the fractal dimension is identified with the number of GMs.  Section~\ref{sum} is devoted to a summary.

\section{The model Hamiltonian}~\label{mh}
The model Hamiltonian under consideration consists of dimer and trimer projection operators, taking the form~\cite{dtmodel}
\begin{align}
H = -\sum_{i}\bigl( \cos \theta \,\, D(i) + \sin \theta \, \, T(i) \bigr).
\label{model}
\end{align}
In an obvious notation, the operators $D(i)$ and $T(i)$ are proportional to the dimer and trimer projection operators.
The parameter $\theta$ represents the mixing angle quantifying the contributions from the  dimer and trimer interactions. 
The sum over $i$ is from 1 to $L$ for PBCs while for OBCs the sum is from 1 to $L-1$ for the dimer term and from 1 to $L-2$ for the trimer term.
Using the spin-1 operator $\v S_i$ at each lattice site, $\v S_{ij} = \v S_i + \v S_j$ for a pair of adjacent sites ($j=i+1$), and $\v S_{ijk} = \v S_i + \v S_j+ \v S_k$ for a triplet of adjacent sites ($k=i+2$), the dimer and trimer projection operators can be expressed as
\begin{align}
{\mathcal P}_D(i) & =  \frac{1}{12} \left( \v S_{ij}^2 -2 \right) \left(\v  S_{ij}^2 -6\right)=  \frac{1}{3} \left( \v S_i \cdot \v S_{j}\right)^2 -\frac{1}{3} , \nn
{\mathcal P}_T(i) & =  - \frac{1}{144}  \left(\v S_{ijk}^2 - 2\right) \left(\v S_{ijk}^2 - 6 \right) \left(\v S_{ijk}^2 - 12\right). \label{eq:dimer-and-trimer-projector}
\end{align}
Each projection operator gives $+1$ for the spin singlets, and $0$ for all other spin multiplets.  The projectors are related to $D(i)$ and $T(i)$ in Hamiltonian (\ref{model}) via 
\begin{align*}
D(i) = 3 \, {\mathcal P}_D(i), ~~~
T(i) = 6 \, {\mathcal P}_T(i).
\label{eq:DT_operators}
\end{align*}

\begin{figure}[t]
	\includegraphics[width=0.3\textwidth]{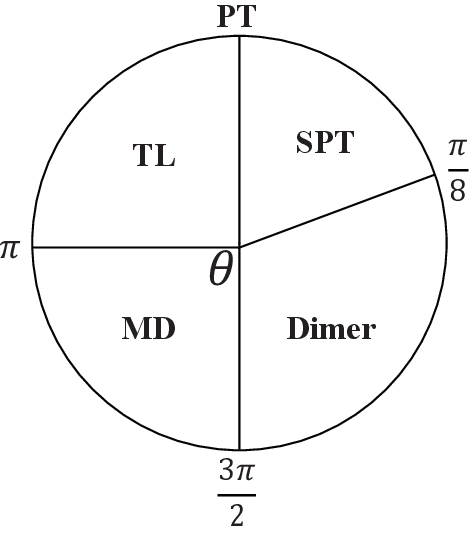}
	\caption{Ground state phase diagram for the quantum spin-1 system with competing dimer and trimer interactions (adapted from Refs.~\cite{dtmodel} and~\cite{dtmodel2}). The angle $\theta$ represents the mixing angle quantifying the contributions from the  dimer and trimer interactions in the model Hamiltonian (\ref{model}). The indicated phases are discussed in Section~\ref{mh}.}
	\label{fig:PD}
\end{figure}

In Figure \ref{fig:PD}, we show the ground state phase diagram of the model Hamiltonian~(\ref{model}) with competing dimer and trimer interactions, adapted from Refs.~\cite{dtmodel} and~\cite{dtmodel2}. Here abbreviations are exploited to label the distinct phases. Specifically, MD stands for macroscopically degenerate phase, SPT for symmetry-protected topological phase, and  TL for trimer liquid phase. The dimerized phase is topologically trivial and gapped, with ground states breaking the  symmetry under the one-site translation operation. In contrast, the Haldane (SPT) phase is topologically non-trivial and translation-invariant. The TL phase is critical with central charge $c=2$. The ground state degeneracy in the MD phase is exponential as a function of the system size $L$.

\section{Ground state subspaces and degeneracies}~\label{gss}
Macroscopically degenerate ground states are present in the region $\pi \le \theta \le 3 \pi/2$, including the MD phase ($\pi < \theta < 3 \pi/2$) and the two endpoints, i.e., the dimer point ($\theta = \pi$) and the trimer point ($\theta = 3\pi/2$). As demonstrated in Ref.~\cite{dtmodel},  the ground-state degeneracies $\Omega_L^{\rm OBC/PBC}$ under OBCs and PBCs scale exponentially for large $L$. 
Specifically, at $\theta=\pi$, $\Omega_L^{\rm OBC/PBC} \sim  \varphi^L$, where $\varphi =(1+\sqrt{5})/2$ is the golden ratio. 
The ground state degeneracies $\Omega_L^{\rm OBC/PBC}$ asymptotically thus become the well-known self-similar golden spiral~\cite{golden}.  At $\theta=3 \pi/2$, $\Omega_L^{\rm OBC/PBC} \sim (2.879)^L$.
For $\pi < \theta < 3 \pi/2$, $\Omega_L^{\rm OBC/PBC} \sim (2.412)^L$. 
It follows in both cases that the ground state degeneracies $\Omega_L^{\rm OBC/PBC}$ asymptotically become a self-similar logarithmic spiral.
Explicit formulas for the ground-state degeneracies $\Omega_L^{\rm OBC/PBC}$ are given in Section D.

\subsection{Highly degenerate ground states at the dimer point}
At the dimer point ($\theta=\pi$) the Hamiltonian (\ref{model}) is identical to the staggered spin-1 ferromagnetic biquadratic model, a special point in the widely studied ${\rm SU}(2)$ spin-1 bilinear-biquadratic model~\cite{AKLT, Sutherland, TB, barber, Chubukov, Fath,  Kawashima,Ivanov, Rizzi, Lauchli, Porras, Romero, ronny, Rakov, Sierra, daibb}. 
The staggered spin-1 ferromagnetic biquadratic model has been systematically investigated recently in Ref.~\cite{golden}.
At this point, the ground state degeneracies $\Omega_L^{\rm OBC/PBC}$ are characterized by the Fibonacci-Lucas sequences~\cite{spins,saleur,dtmodel,moudgalya}, with non-zero residual entropy. 
This implies that the ground state degeneracies  are asymptotically the golden spiral.
In fact, the highly degenerate and highly entangled ground states, arising from a SSB pattern from ${\rm SU}(3)$ to ${\rm U}(1)\times {\rm U}(1)$ with two type-B GMs,
are scale invariant, but not conformally invariant. 
The entanglement entropy scales logarithmically  with the block size $n$  in the thermodynamic limit $L \rightarrow \infty$, with the prefactor being half the number of type-B GMs. The latter in turn is identified to be the fractal dimension.

We emphasize that sequences of degenerate ground states are generated {\it not only} from the (unique) highest weight state, {\it but also} from generalized highest weight states. As argued, the necessity for introducing the notion of a generalized highest weight state lies in the fact that, for the staggered ${\rm SU}(3)$ ferromagnetic biquadratic model, or equivalently the dimer point ($\theta=\pi$), the ground state degeneracies are exponential as a function of the system size $L$.

For later use, we introduce some terminology and notation. The highest weight state is denoted by $\vert{\rm hws}\rangle_q$, with the period $q$ being  1 or 2, and a  generalized highest weight state is denoted by
$\vert{\rm ghws}\rangle_p$, with the period $p$ being an integer, not less than two. Actually, if the symmetry group is staggered, then the highest weight state  $\vert{\rm hws}\rangle_q$ might be staggered, so the period $q$ may be one or two, depending on the choices of the Cartan generators. However, if the symmetry group is uniform, then the period $q$ of the highest weight state $\vert{\rm hws}\rangle_q$ is always one. Note that the period $q$ or $p$ may be regarded as the size of the unit cell for an exact matrix product state (MPS) representation of highly degenerate ground states~\cite{exactmps}. From now on, a convention is adopted for the subscripts $q$ and $p$ that we only use $p$ to label the norms and the states generated from the highest weight state and the generalized highest weight states, since the symmetry group is uniform at the trimer point ($\theta=3\pi/2$) and  in the dimer-trimer regime ($\pi<\theta<3\pi/2$). As a result, $p$ is an integer not less than one.

We emphasize that, in contrast to the highest weight state, a generalized highest state is not unique. Indeed, it normally depends on what type of boundary conditions are adopted.
Formally, the $m$-th generalized highest weight state $|{\rm ghws}\rangle_m$ is defined recursively as $|{\rm ghws}\rangle_m\notin V_0\oplus V_1\oplus... V_{m-1}$, $E_1|{\rm ghws}\rangle_m=0\mod(V_0\oplus V_1\oplus... V_{m-1})$, and $E_2|{\rm ghws}\rangle_m=0\mod(V_0\oplus V_1\oplus... V_{m-1})$~\cite{golden}.
Here $V_0$ denotes the subspace spanned by the degenerate ground states generated from the highest weight state, whereas $V_\mu$ denotes the subspace spanned by the degenerate ground states generated from the $\mu$-th generalized highest state, where  $\mu=1,2, \ldots, m$.
This means that, even though $|{\rm ghws}\rangle_m$ are linearly independent to the states in the subspace  $V_0\oplus V_1\oplus... V_{m-1}$, all of the states generated from $|{\rm ghws}\rangle_m$, with $E_1^{M_1}E_2^{M_2}|{\rm ghws}\rangle_m$ ($M_1\geq0$, $M_2\geq 0$ and $M_1+M_2\geq1$), are linearly dependent to the states in the subspace $V_0\oplus V_1\oplus... V_{m-1}$.
Once a complete set of the generalized highest weight states are determined, the subscript $m$ is dropped for brevity. Instead, the subscript $p$, which denotes the period, is used to label a generalized highest weight state.

\subsection{Highly degenerate ground states at the trimer point}

The Hamiltonian (\ref{model}) at the trimer point ($\theta=3\pi/2$) possesses the uniform  symmetry group ${\rm SU}(3)$, with two (commuting) Cartan generators $H_1$ and $H_2$.
For each of these generators there exists a conjugate pair of a raising operator and a lowering operator: $E_1$ and $F_1$, and  $E_2$ and $F_2$.
At the trimer point, sequences of degenerate ground states are generated from the repeated action of $F_1$ and  $F_2$, combining with a  symmetry operation $\mathscr{S}$ from the symmetric group $S_3$, which consists of the permutation operations acting on the three local orthonormal basis states $+$, $0$ and $-$ at each lattice site, on both the highest weight state and a generalized highest weight state.
An additional feature is that not all of the generalized highest weight states are (unentangled) factorized.
As an illustration,  for $L=5$, a generalized highest weight state, which is an eigenvector of $S^z$ with eigenvalue two,
takes the form
\begin{align*}
	|\rm{ghws}\rangle_5=	| + \rangle 
        ( | 0+-++ \rangle +| \! +   0   -++ \rangle 
            +| \!+ 0 +  - + \rangle ).
\end{align*}

For an arbitrary system size $L$, we choose a few generalized highest weight states $\vert{\rm ghws}\rangle_p$: $\vert\otimes_{k=1}^{L/2} \{+0\}_{\;k}\rangle$, $\vert\otimes_{k=1}^{L/3}  \{+0+\}_{\;k}\rangle$,   $\vert\otimes_{k=1}^{L/4}  \{+0\!+\!+\}_{\;k}\rangle$, $\vert\otimes_{k=1}^{L/5}  \{+0\!+\!+0\}_{\;k}\rangle$, $\vert\otimes_{k=1}^{L/6}  \{+0\!+\!+\!+\!+\}_{\;k}\rangle$, with corresponding periods $p=2,3,4,5$ and $6$.

A sequence of degenerate ground states $|L,M\rangle_p$ are generated from the repeated action of the lowering operators $F_1$ and $F_2$ on the highest weight state $|{\rm hws}\rangle_p$ or a generalized highest weight state $|{\rm ghws}\rangle_p$: $|L,M_1,M_2\rangle_p=1/Z_p(L,M_1,M_2)F_1^{M_1}F_2^{M_2}|{\rm ghws}\rangle_p$,
where   $Z_p(L,M_1,M_2)$ are introduced to ensure that $|L,M_1,M_2\rangle_p$ are normalized.

The highest weight state with period $p=1$ is $|{\rm hws}\rangle_1=\vert\otimes_{k=1}^{L} \{+\}_{\;k}\rangle$. Hence  $Z_1(L,M_1,M_2)$ is introduced to ensure that $|L,M_1,M_2\rangle_1$ is normalized,
\begin{equation}
	Z_1(L,M_1,M_2)=M_1!M_2!\sqrt{C_{L}^{M_1}C_{L-M_1}^{M_2}},\label{zm1m2}
\end{equation}
with $C_L^{M_1}$ and $C_{L-M_1}^{M_2}$  binomial coefficients.
A generalized highest weight state with period $p=3$ is $|{\rm ghws}\rangle_3=\vert\otimes_{k=1}^{L/3} \{++0\}_{\;k}\rangle$. Hence  we have
\begin{equation}
	Z_3(L,M_1,M_2)=M_1!M_2!\sqrt{C_{2L/3}^{M_1}C_{2L/3-M_1}^{M_2}}.\label{zm1m2p3}
\end{equation}

For an arbitrary period $p$, a sequence of degenerate ground states $|L,M_1,M_2\rangle_p$ are generated from the repeated action of the lowering operators $F_1$ and $F_2$  on 
$|{\rm ghws}\rangle_p$:
\begin{equation}
	\vert L,M_1,M_2\rangle_p=\frac{1}{Z_p(L,M_1,M_2)}F_1^{M_1}F_2^{M_2}\vert{\rm ghws}\rangle_p,
	\label{lm1m2p}
\end{equation}
where $M_1=0$, \ldots, $\gamma L/p$, $M_2=0$, \ldots, $\gamma L/p$, with $\gamma$ denoting the total number of sites in the local orthonormal state $|+\rangle$ in a unit cell for a generalized highest weight state, and $Z_p(L,M_1,M_2)$ is introduced to ensure that $|L,M_1,M_2\rangle_p$ is normalized. Hence  we have
\begin{equation}
	Z_p(L,M_1,M_2)=M_1!M_2!\sqrt{C_{\gamma L/p}^{M_1}C_{\gamma L/p-M_1}^{M_2}}.\label{zm1m2p}
\end{equation}

\subsection{Highly degenerate ground states in the dimer-trimer regime}

For $\pi < \theta < 3 \pi/2$, Hamiltonian (\ref{model}) possesses the uniform symmetry group ${\rm SU}(2)$ generated by $S^x=\sum_i S_i^x$, $S^y=\sum_i S_i^y$ and $S^z=\sum_i S_i^z$, with one Cartan generator $S^z$, one lowering operator $S^-$ and one raising operator $S^+$, defined as $S^{\pm} = (S^x \pm i S^y)/\sqrt {2}$. Note that we stick to the conventional notations for generators of the uniform symmetry group ${\rm SU}(2)$.

The highly degenerate ground states may be generated from the repeated action of the lowering operator $S^-$ on both the highest weight state and generalized highest weight states, combining with the $Z_2$ symmetry operation $I_b$ for OBC and $\tau$ for PBC.
An additional feature is that not all the generalized highest weight states are (unentangled) factorized.
As an illustration, for $L=4$, a generalized highest weight state, which is an eigenvector of $S^z$ with eigenvalue one, takes the form
\begin{align*}
&|\Psi\rangle=2(|-0++\rangle-|++0-\rangle)+(|0-++\rangle-|++-0\rangle) \\
&-(|0+-+\rangle-|+-+0\rangle)-(|-++0\rangle-|0++-\rangle).
\end{align*}

For an arbitrary system size $L$, we choose a few generalized highest weight states $\vert{\rm ghws}\rangle_p$: $\vert\otimes_{k=1}^{L/2} \{+0\}_{\;k}\rangle$, $\vert\otimes_{k=1}^{L/3}  \{+0+\}_{\;k}\rangle$,   $\vert\otimes_{k=1}^{L/4}  \{+0++\}_{\;k}\rangle$, $\vert\otimes_{k=1}^{L/5}  \{+0++0\}_{\;k}\rangle$, $\vert\otimes_{k=1}^{L/6}  \{+0++++\}_{\;k}\rangle$, with period $p=2,3,4,5$ and $6$, respectively. We remark that the chosen generalized highest weight states in the dimer-trimer regime ($\pi<\theta<3\pi/2$) are identical to those at the trimer point ($\theta = \pi$). However, we emphasize that this is {\it only} a choice suitable to a further finite system-size scaling analysis of the entanglement entropy. Not all of the generalized highest weight states are actually identical for the two SSB patterns.

A sequence of degenerate ground states   $|L,M\rangle_p$ are generated from the repeated action of the lowering operators $S^-$ on the highest weight state $|{\rm hws}\rangle_p$ or a generalized highest weight state $|{\rm ghws}\rangle_p$: $|L,M\rangle_p=1/{Z_p(L,M)}(S^-)^{M}|{\rm ghws}\rangle_p$,
where $Z_p(L,M)$ are introduced to ensure that  $|L,M\rangle_p$ are normalized.

\begin{widetext}
	The highest weight state with period $p=1$ is $|{\rm hws}\rangle_1=\vert\otimes_{k=1}^{L} \{+\}_{\;k}\rangle$. Hence  we have
	\begin{equation}
		Z_1(L,M)=M!\sqrt{C_{L}^{M}}\label{zmsu21}.
	\end{equation}	
A generalized highest weight state with period $p=2$ is $|{\rm ghws}\rangle_2=\vert\otimes_{k=1}^{L/2} \{+0\}_{\;k}\rangle$. Hence  we have
	\begin{equation}
		Z_2(L,M)=M!\sqrt{\sum_{j=0}^{{\rm min}(L/2,M)}\frac{1}{2^{M-j}}C_{L/2}^j{\sum}'_{N_{+},N_{0},\;N_{-}}2^{N_0-L/2+j}C_{L/2}^{N_--j}C_{L/2-N_{-}+j}^{N_0-L/2+j}}.\label{zmsu22}
	\end{equation}
For an arbitrary period $p$, 
\begin{equation}
	\vert L,M\rangle_p=\frac{1}{Z_p(L,M)} (S^-)^M \vert{\rm ghws}\rangle_p,
	\label{lm1m2p}
\end{equation}
where $M=0$, \ldots, $\gamma L/p$, with $\gamma$ denoting the total number of sites in the local orthonormal state $|+\rangle$ in a unit cell for a generalized highest weight state, and $Z_p(L,M)$ is introduced to ensure that $|L,M\rangle_p$ is normalized.
In general, 
\begin{equation}
	Z_p(L,M)=M!\sqrt{\sum_{j=0}^{{\rm min}((1-\gamma/p) L,M)}\frac{1}{2^{M-j}}C_{(1-\gamma/p) L}^j{\sum}'_{N_{+},N_{0},\;N_{-}}2^{N_0-(1-\gamma/p) L+j}C_{\gamma L/p}^{N_--j}C_{\gamma L/p-N_{-}+j}^{N_0-\gamma L/p+j}} \,.
	\label{zplmsu2}
\end{equation}
\end{widetext}

We may choose $\vert\otimes_{k=1}^{L/2} \{0\rangle(|+\rangle+|-\rangle)\}_{\;k}\rangle$ as an alternative generalized highest weight state, with the period $p$ being two. A sequence of degenerate ground states $|L,M\rangle_2$, are generated from the repeated action of the lowering operators $S^-$ on the factorized state $\vert\otimes_{k=1}^{L/2} \{0\rangle(|+\rangle+|-\rangle)\}_{\;k}\rangle$:
\begin{equation}
	|L,M\rangle_2=\frac{1}{Z_2(L,M)}(S^-)^{M}|0\rangle(|+\rangle+|-\rangle) ...|0\rangle(|+\rangle+|-\rangle).
	\label{lmsu22}
\end{equation}

Needless to say, many other choices are possible for a generalized highest weight state $|{\rm ghws} \rangle_p$ with the period $p$. However, we restrict ourselves to a few of them to demonstrate how highly degenerate ground states are generated from generalized highest weight states at the trimer point and in the dimer-trimer regime.

\subsection{Ground state degeneracies in the macroscopically degenerate regime}

Here we focus on the regime $\pi\leq\theta\leq3\pi/2$.
Let us first start with OBCs. At the dimer point $\theta = \pi$, the  ground state degeneracy $\Omega_L^{\rm OBC}$ follows from the recursive relation
\begin{equation}
	\Omega_L^{\rm OBC} = 3 \, \Omega_{L-1}^{\rm OBC} - \Omega_{L-2}^{\rm OBC},
\end{equation}
with $\Omega_1^{\rm OBC}=3$ and $\Omega_2^{\rm OBC}=8$.
Hence we have
\begin{equation}
	\Omega_L^{\rm OBC}= \frac{(3+\sqrt 5)^{L+1} - (3-\sqrt 5)^{L+1}}{2^{L+1} \sqrt{5}} .
\end{equation}
Here $(3\pm\sqrt 5)/2$ are the two roots of the characteristic equation $x^2-3x+1=0$.

At the trimer point $\theta = 3\pi/2$, the ground state degeneracy $\Omega_L^{\rm OBC}$ follows from the recursive relation
\begin{equation}
	\Omega_L^{\rm OBC}=3 \, \Omega_{L-1}^{\rm OBC}-\Omega_{L-3}^{\rm OBC},
\end{equation}
with $\Omega_1^{\rm OBC}=3$, $\Omega_2^{\rm OBC}=9$ and $\Omega_3^{\rm OBC}=26$.
Hence we have
\begin{equation}
	\Omega_L^{\rm OBC}=\frac{c_1^{L+1}-c_2^{L+2}+c_2c_3^{L+2}}{3(c_1-2)}.
\end{equation}
Here $c_1$, $c_2$ and $c_3$ are the roots of the characteristic equation $x^3-3x^2+1=0$. These are
$c_1= 1-2\cos(8\pi/9)$, $c_2=1-2\cos(4\pi/9)$ and $c_3=1-2\cos(2\pi/9)$.

For the dimer-trimer regime $\pi <\theta < 3\pi/2$, the ground state degeneracy $\Omega_L^{\rm OBC}$ follows from the recursive relation
\begin{equation}
	\Omega_L^{\rm OBC}=3 \, \Omega_{L-1}^{\rm OBC}-\Omega_{L-2}^{\rm OBC}-\Omega_{L-3}^{\rm OBC},
\end{equation}
with $\Omega_1^{\rm OBC}=3$, $\Omega_2^{\rm OBC}=8$ and $\Omega_3^{\rm OBC}=20$.
Hence we have
\begin{equation}
	\Omega_L^{\rm OBC}=\frac{(1+\sqrt{2})^{L+2}+(1-\sqrt{2})^{L+2}-2}{4}.
\end{equation}
Here  $1\pm\sqrt{2}$ and 1 are the roots of the characteristic equation $x^3-3x^2+x+1=0$.

\begin{table*}[htb]
	\begin{tabular}{c|cccccccc|c}
		\hline \hline
		$L$ & \,~3~\, & 4 & 5 &6 & 7 & 8 & 9 & 10 & OEIS  \\ \hline
		$\theta=\pi$ & ~~21 ~ & ~55~ & ~144 ~ & ~377~ & ~987 ~ & ~2584~ & ~6765~ & ~17711~ & A001906   \\
		$\pi < \theta < 3  \pi/2$ & \,~20~\, & ~49~ & ~119~ & ~288~ & ~696~ & ~1681~ & ~4059~ & ~9800~ & A048739 \\
		$\theta=3 \pi/2$ & \,~26~\, & ~75~ & ~216~ & ~622~ & ~1791~ & ~5157~ & ~14849~ & ~42756~  &
		A076264 \\
		\hline \hline
	\end{tabular}
	\caption{Ground state degeneracies  for the quantum spin-1 dimer-trimer model with OBCs for increasing system size $L$. The rightmost column shows the corresponding integer sequences in the On-Line Encyclopedia of Integer Sequences (OEIS).}
	\label{tab:OBC}
\end{table*}

\begin{table*}[htb]
	\begin{tabular}{c|cccccccc|c}
		\hline \hline
		$L$ & \,~3~\, & 4 & 5 &6 & 7 & 8 & 9 & 10 & OEIS \\ \hline
		$\theta=\pi$ & ~~18 ~ & ~47~ & ~123 ~ & ~322~ & ~843 ~ & ~2207~ & ~5778~ & ~15127~ & A005248  \\
		$\pi < \theta < 3\pi/2$ & \,~17~\, & ~41~ & ~83~ & ~209~ & ~479~ & ~1169~ & ~2787~ & ~6745~ & $-$  \\
		$\theta=3\pi/2$ & \,~26~\, & ~72~ & ~198~ & ~570~ & ~1641~ & ~4725~ & ~13605~ & ~39174~ & $A215885$ \\
		\hline \hline
	\end{tabular}
	\caption{Ground state degeneracies  for the quantum spin-1 dimer-trimer model with PBCs for increasing system size $L$. The rightmost column shows the corresponding integer sequences in the On-Line Encyclopedia of Integer Sequences (OEIS).}
	\label{tab:PBC}
\end{table*}

Now we turn to PBCs. At the dimer point $\theta = \pi$, the ground state degeneracy $\Omega_L^{\rm PBC}$ follows from
\begin{equation}
	\Omega_L^{\rm PBC}=3\,	\Omega_{L-1}^{\rm OBC}-2 \, \Omega_{L-2}^{\rm OBC}.
\end{equation}	
Hence  $\Omega_L^{\rm PBC}$ takes the form
\begin{align}
	\Omega_L^{\rm PBC}= \frac{(3+\sqrt 5)^L + (3-\sqrt 5)^L}{ 2^L }.
\end{align}

At the trimer point $\theta = 3\pi/2$, the ground state degeneracy $\Omega_L^{\rm PBC}$  follows from
\begin{align}
	\Omega_L^{\rm PBC}=3 \, \Omega_{L-1}^{\rm OBC}-3 \, \Omega_{L-3}^{\rm OBC}.
\end{align}
Hence   $\Omega_L^{\rm PBC}$ ($L\geq5$) takes the form
\begin{equation}
	\Omega_L^{\rm PBC}=c_1^{L}+c_2^{L}+c_3^{L},
\end{equation}
with $c_1$, $c_2$ and $c_3$ as given above for OBCs.

For the dimer-trimer regime $\pi <\theta < 3\pi/2$,  the ground state degeneracy $\Omega_L^{\rm PBC}$ ($L>3$)  follows from
\begin{equation}
	\Omega_L^{\rm PBC}=3 \, \Omega_{L-1}^{\rm OBC}-2 \, \Omega_{L-2}^{\rm OBC}-3\,\Omega_{L-3}^{\rm OBC},
\end{equation}
subject to an odd-even parity effect. Indeed,  $\Omega_L^{\rm PBC}$ takes the form
\begin{equation}
	\Omega_L^{\rm PBC}=(1+\sqrt{2})^{L}+(1-\sqrt{2})^{L}+2L-1
\end{equation}
for $L\in 2\mathbb{Z}_+$, and
\begin{equation}
	\Omega_L^{\rm PBC}=(1+\sqrt{2})^{L}+(1-\sqrt{2})^{L}+1
\end{equation}
for $L \in 2\mathbb{Z}_++1$. Here  $\mathbb{Z}_+$ represents the set of positive integers. We stress that  there is an extra term $2L-2$ in $\Omega_L^{\rm PBC}$ for $L\in 2\mathbb{Z}_+$. This extra contribution to the ground state degeneracy under PBCs originates from a parity effect for the model. 
Actually, the staggered ${\rm SU}(3)$ biquadratic model at the dimer point  ($\theta=\pi$) also exhibits a parity effect under PBCs. That is, if $L$ is even, then
the symmetry group is the staggered ${\rm SU}(3)$ symmetry group, with two type-B GMs.
On the other hand, if $L$ is odd, then
the symmetry group is the uniform ${\rm SU}(2)$ symmetry group, with one type-B GM~\cite{golden}. In fact, if the angle $\theta$ varies from the dimer-trimer regime to the dimer point, the symmetry group suddenly changes from ${\rm SU}(2)$ to ${\rm SU}(3)$ if $L$ is even, but remains the same ${\rm SU}(2)$ if $L$ is odd. In this sense, one may regard the extra $2L-2$ linearly independent ground states in the dimer-trimer regime as a remnant of one extra GM at the dimer point, if $L$ is even. Here we remark that the number of degenerate ground states to support one type-B GM is linear in $L$, as one sees from the ${\rm SU}(2)$ ferromagnetic Heisenberg model in one spatial dimension, given that one GM requires one ${\rm SU}(2)$ subgroup.
Although the symmetry groups are uniform in the dimer-trimer regime ($\pi <\theta < 3\pi/2$)  and at the trimer point ($\theta=3\pi/2$), an emergent local symmetry operation exists, as discussed in Subsection $E$. Therefore the parity effect emerges in the dimer-trimer regime as a result of such an emergent local symmetry operation.

For small system sizes, e.g., $L=3$, $4$ and $5$, we may figure out all of the generalized highest weight states, in addition to the unique highest weight state, under both OBCs and PBCs. Acting the lowering operator $S^-$ on the highest weight state and the generalized highest weight states, we are able to produce all linearly independent degenerate ground states.  Meanwhile, the explicit expressions for the ground state degeneracies $\Omega_L^{\rm OBC/PBC}$ for the three SSB patterns reproduce the ground state degeneracies obtained from exact diagonalization, with system size up to $L=10$ (cf. Table~\ref{tab:OBC} and Table~\ref{tab:PBC}). We remark that such formulae are reminiscent of the Binet formula which expresses an integer in terms of an irrational number~\cite{binet}. In this way the connection between the Fibonacci-Lucas sequences and the ground state degeneracies under OBCs and PBCs at the dimer point~\cite{golden} is extended to the trimer point and the dimer-trimer regime, with the Fibonacci-Lucas sequences replaced by other sequences of integers. As already pointed out in Ref.~\cite{dtmodel}, some of these sequences of integers have appeared in the On-Line Encyclopedia of Integer Sequences (OEIS).

An additional consequence one may draw from the fact that the  ground state degeneracies $\Omega_L^{\rm OBC/PBC}$ are exponential is that the residual entropy $S_{\!r}$ is non-zero. Specifically, we have
$S_{\!r} =\ln((3+\sqrt 5)/2) $ at the dimer point $\theta = \pi$, $S_{\!r} = \ln(1-2\cos(8\pi/9)) $ at the trimer point $\theta = 3\pi/2$, and $S_{\!r} = \ln(1+\sqrt{2})$ in the dimer-trimer regime $\pi <\theta < 3\pi/2$. Physically, the residual entropy measures the disorder present in a unit cell of an MPS representation for a highly degenerate ground state generated from a generalized highest weight state (for a detailed account of an exact MPS representation for the spin-1 ferromagnetic biquadratic model, cf. Ref.~\cite{exactmps}).

\subsection{The origin of the ground state degeneracies as an exponential function of the system size}

At the dimer point $\theta = \pi$, the symmetry group ${\rm SU}(3)$ is staggered. Thus, some generators do not commute with the one-site translation operation $\tau$ under PBCs or the bond-centered/site-centered inversion $I_{b/s}$ for even/odd $L$ under OBCs. Here $I_{b}$ or $I_{s}$ is defined as an inversion operation with respect to the bond or site in the middle of the one-dimensional lattice. As already discussed in Ref.~\cite{golden}, it is this staggered nature that accounts for the exponential ground state degeneracies with the system size $L$. However, this interpretation does not work in the dimer-trimer regime $\pi <\theta < 3\pi/2$ and at the trimer point $\theta = 3\pi/2$, since the symmetry groups are uniform.
This implies that a proper approach to account for the exponential ground state degeneracies $\Omega_L^{\rm OBC/PBC}$ with the system size $L$ remains to be developed, which should be able to tackle the questions why the ground state degeneracies $\Omega_L^{\rm OBC/PBC}$ are exponential as a function of the system size $L$
and why the ground state degeneracies $\Omega_L^{\rm OBC/PBC}$ depend on what types of the boundary conditions adopted for the three SSB patterns under investigation. Indeed, a careful examination reveals an {\it emergent} local symmetry operation that is {\it tailored} to a specific degenerate ground state in the dimer-trimer regime $\pi <\theta < 3\pi/2$ and at the trimer point $\theta = 3\pi/2$.

For this purpose, we resort to a simple mathematical lemma. Suppose there exists a (local) unitary operation $g$ that does not commute with the Hamiltonian $H$, i.e., $[H,g] \neq 0$. Denote the commutator between  $H$ and $g$ as $V$: $V=Hg -gH$.
For a specific ground state $|\psi_0\rangle$, if $V|\psi_0\rangle=0$ and $|\langle\psi_0| g|\psi_0\rangle|\neq 1$, then the action of $g$ on $|\psi_0\rangle$ yields a degenerate ground state.
The proof simply follows from the observation that $Hg|\psi_0\rangle=gH|\psi_0\rangle=E_gg|\psi_0\rangle$.
Given $|\langle\psi_0| g|\psi_0\rangle|\neq 1$,  $g|\psi_0\rangle$ must be a degenerate ground state.  In this sense, a unitary operation $g$ is said to be an emergent (local) symmetry operation tailored to a specific degenerate ground state $|\psi_0\rangle$. Normally, $g$ itself generates an emergent discrete symmetry group $G_d$ tailored to a specific ground state  $|\psi_0\rangle$.  If a unitary operation is an emergent symmetry operation in the above sense for any degenerate ground state, then it is said to be an emergent symmetry operation (tailored to the ground state subspace). In fact, there are a series of  emergent (local) symmetry operations tailored to different degenerate ground states, labeled by the eigenvalues of $S^z$, for the three SSB patterns. In particular, their presence is essential to the exponential ground state degeneracies under OBCs and PBCs in the dimer-trimer regime $\pi <\theta < 3\pi/2$ and at the trimer point $\theta = 3\pi/2$.

To  proceed, we remark that the model Hamiltonian (\ref{model}) commutes with the one-site translation operation $\tau$ under PBCs. 
Meanwhile the generator $\sigma$, defined as $\sigma \equiv P_{12}P_{23} \ldots P_{L-1L}$, of a cyclic permutation group $Z_L$ is an emergent symmetry operation under OBCs, where $P_{ii+1}$ is a cyclic permutation acting on the two adjacent sites $i$ and $i+1$.
Actually the action of the one-site translation operation on a given ground state under PBCs is identical to that of
the generator $\sigma$ under OBCs, if this ground state remains to be the same as the boundary conditions varies from OBCs to PBCs.

For our purpose it is proper to start from the highest weight state $\vert\otimes_{k=1}^{L} \{+\}_{\;k}\rangle$, an eigenvector of $S^z$ with the eigenvalue $L$.  If $|\psi_0\rangle$ is chosen to be $S^-\vert\otimes_{k=1}^{L} \{+\}_{\;k}\rangle$, an eigenvector of $S^z$ with the eigenvalue being $L-1$, then we consider a set of local unitary operations $g_j$ (independent of the boundary conditions), defined as $\exp (i \pi \Sigma_j)$, with $\Sigma_j =S_j^z$ ($j=1$, \ldots, $L$). As follows from the above lemma, we are led to a set of degenerate ground states  $g_j|\psi_0\rangle$,
which take the form: $\sum _{i=1}^L (1-2\delta_{i\;j}) |+_1\ldots +_{i-1} 0_i +_{i+1} \ldots +_L \rangle$, up to a multiplicative constant.
We remark that each of $g_j$'s generates an emergent discrete symmetry group $Z_2$ tailored to $|\psi_0\rangle$, because $g_j^2|\psi_0\rangle= |\psi_0\rangle$.
Thus we are led to degenerate ground states: $\sigma^i |0_1+_2 \ldots + \rangle$ ($i=1,2,\ldots,L$) under OBCs, which are identical to $\tau^i |0_1+_2 \ldots + \rangle$  ($i=1,2,\ldots,L$)  under PBCs. Note that $\sigma^i$ and $\tau^i$ denote the $i$-th power of $\sigma$ and $\tau$, respectively.
Actually, they constitute a bunch of generalized highest weight states, valid for the three SSB patterns.
In particular, the argument also applies to the dimer point $\theta = \pi$.  We remark that an explicit form of the commutator $V$ between the Hamiltonian and a local unitary operation is given in Appendix B.

Now we turn to a generalized highest weight state as an eigenvector of $S^z$ with the eigenvalue $L-2$. If $|\psi_0\rangle$ is chosen to be $S^-\vert 0_1+_2 \ldots +_L\rangle$, then we consider a local unitary operation $g$, defined as $\exp (i \pi \Sigma)$, with  $\Sigma=S_1^z+S_2^z$ under OBCs and $\Sigma=S_1^z+S_2^z+S_L^z$ under PBCs.
As follows from the lemma, we are led to a degenerate ground state  $g|\psi_0\rangle$, which takes the form: $\vert -_1+_2\ldots+_L\rangle+\vert 0_10_2\ldots+_L\rangle- \sum _{i=3}^{L} |0_1+_2\ldots +_{i-1} 0_i +_{i+1} \ldots +_L \rangle$ under OBCs and
$-\vert -_1+_2\ldots+_L\rangle-\vert 0_10_2\ldots+_L\rangle +\sum _{i=3}^{L-1} |0_1+_2\ldots +_{i-1} 0_i +_{i+1} \ldots +_L \rangle-\vert 0_1+_2\ldots+_{L-1}0_L\rangle$ under PBCs, up to a multiplicative constant. Again, $g$ generates an emergent discrete  symmetry group $Z_2$ tailored to $|\psi_0\rangle$, because $g^2|\psi_0\rangle= |\psi_0\rangle$.
Thus we are led to degenerate ground states: $\vert -_1+_2\ldots+_L\rangle+\vert 0_10_2\ldots+_L\rangle$ and $\sum _{i=3}^{L} |0_1+_2\ldots +_{i-1} 0_i +_{i+1} \ldots +_L\rangle$ under OBCs and $\vert -_1+_2\ldots+_L\rangle+\vert 0_10_2\ldots+_L\rangle +\vert 0_1+_2\ldots+_{L-1}0_L\rangle$ and $\sum _{i=3}^{L-1} |0_1+_2\ldots +_{i-1} 0_i +_{i+1} \ldots +_L \rangle$ under PBCs. Afterwards,
if $|\psi_0\rangle$ is chosen to be $\sum _{i=3}^L |0_1+_2\ldots +_{i-1} 0_i +_{i+1} \ldots +_L \rangle$ under OBCs, then we consider
a set of local unitary operations $g_j$, defined as $\exp (i \pi \Sigma_j)$, with $\Sigma_j =S_j^z$ ($j=3$, \ldots, $L$). Therefore, we are led to a set of degenerate ground states  $g_j|\psi_0\rangle$,
which take the form: $\sum _{i=3}^{L} (1-2\delta_{i\;j}) |0_1+_2+_1\ldots +_{i-1} 0_i +_{i+1} \ldots +_L \rangle$, up to a multiplicative constant.
Again, each of $g_j$'s generates an emergent discrete symmetry group $Z_2$ tailored to $|\psi_0\rangle$, because $g_j^2|\psi_0\rangle= |\psi_0\rangle$.
Thus we are led to degenerate ground states: $|0_1+_2 \ldots +_{i-1}0_i+_{i+1}\ldots\rangle$ ($ i=3,4,\ldots,L$).
Similarly, if $|\psi_0\rangle$ is chosen to be $\sum _{i=3}^{L-1} |0_1+_2\ldots +_{i-1} 0_i +_{i+1} \ldots +_L \rangle$ under PBCs, then we consider
a set of local unitary operations $g_j$, defined as $\exp (i \pi \Sigma_j)$, with $\Sigma_j = S_j^z$ ($j=3$, \ldots, $L-1$). Therefore, we are led to a set of degenerate ground states  $g_j|\psi_0\rangle$,
which take the form $\sum _{i=3}^{L-1} (1-2\delta_{i\;j}) |0_1+_2+_1\ldots +_{i-1} 0_i +_{i+1} \ldots +_L \rangle$, up to a multiplicative constant.
Again, each of $g_j$'s generates an emergent discrete symmetry group $Z_2$ tailored to $|\psi_0\rangle$, because $g_j^2|\psi_0\rangle= |\psi_0\rangle$.
Thus we are led to degenerate ground states: $|0_1+_2 \ldots +_{i-1}0_i+_{i+1}\ldots\rangle$ ($ i=3,4,\ldots,L-1$).
This vividly illustrates why the ground state degeneracies depend on what types of boundary conditions are adopted, since different boundary conditions lead to different generalized highest weight states.

One may extend this construction to other emergent (local) symmetry operations tailored to other specific degenerate ground states, which are eigenvectors of $S^z$ with the eigenvalues less than $L-2$.  Hence it is the presence of  emergent local symmetry operations tailored to a specific degenerate ground state that is responsible for the fact that the ground state degeneracies $\Omega_L^{\rm OBC/PBC}$ are exponential as a function of the system size $L$. Here we stress that the choices of emergent (local) symmetry operations tailored to specific degenerate ground states
are not unique. In fact,  there are other choices that produce the same set of generalized highest weight states. Therefore, a systematic investigation into the role of emergent (local) symmetry operations is necessary for a full understanding of the inherent structure underlying the ground state space, which is deferred to a forthcoming article. Indeed, the powerfulness of  emergent (local) symmetry operations remains to be exposed.

There is another remarkable difference between the SSB pattern at the dimer point and the SSB patterns in the dimer-trimer regime and at the trimer point. At the dimer point, all generalized highest weight states are factorized~\cite{golden}. In contrast, at least local (short-ranged) entanglement is present in some generalized highest weight states in the dimer-trimer regime and at the trimer point, as already shown in  Subsection B and Subsection C for small $L$.

\section{Schmidt decomposition reflecting self-similarities}~\label{svd}
Given that the model at the dimer point $\theta = \pi$ is the staggered ${\rm SU}(3)$ spin-1 ferromagnetic biquadratic model, which has been extensively studied in Ref.~\cite{golden}, we focus here on the dimer-trimer regime $\pi <\theta < 3\pi/2$ and at the trimer point $\theta = 3\pi/2$. We recall that the symmetry groups are uniform ${\rm SU}(2)$ in the dimer-trimer regime and  uniform ${\rm SU}(3)$ at the trimer point.

An exact Schmidt decomposition may be performed for the degenerate ground states  $|L,M_1,M_2\rangle_{p}$ or $|L,M\rangle_{p}$ generated from the highest weight state $\vert{\rm hws}\rangle_1$, or generalized highest weight states $\vert{\rm ghws}\rangle_p$. Here the period $q$ is chosen to be one, since the symmetry group is uniform, and is labeled by $p=1$ according to our convention. Moreover, we only choose the same set of $\vert{\rm hws}\rangle_1$ and $\vert{\rm ghws}\rangle_p$ for the model in the dimer-trimer regime and uniform ${\rm SU}(3)$ at the trimer point.
In this way we have $\vert{\rm hws}\rangle_1 = \vert\otimes_{k=1}^{L} \{+\}_{\;k}\rangle$ and $\vert{\rm ghws}\rangle_2=\vert\otimes_{k=1}^{L/2} \{+0\}_{\;k}\rangle$, $\vert{\rm ghws}\rangle_3=\vert\otimes_{k=1}^{L/3}  \{+0+\}_{\;k}\rangle$,   $\vert{\rm ghws}\rangle_4=\vert\otimes_{k=1}^{L/4}  \{+0++\}_{\;k}\rangle$, $\vert{\rm ghws}\rangle_5=\vert\otimes_{k=1}^{L/5}  \{+0++0\}_{\;k}\rangle$, and $\vert{\rm ghws}\rangle_6=\vert\otimes_{k=1}^{L/6}  \{+0++++\}_{\;k}\rangle$.

At the trimer point $\theta = 3\pi/2$, the degenerate ground states $|L,M_1,M_2\rangle_p$ in Eq.~(\ref{lm1m2p}) admit the exact Schmidt decomposition
\begin{widetext}
	\begin{equation}
		|L,M_1,M_2\rangle_p= \sum\limits_{k_1=0}^{\min(M_1,\gamma n)}\sum\limits_{k_2=0}^{\min(M_2,\gamma n-k_1)}\lambda(L,n,k_1,k_2,M_1,M_2)
		|n,k_1,k_2\rangle_p|L-n,M_1-k_1,M_2-k_2\rangle_p, \label{lm1m2psvd}
	\end{equation}
	where $n$ is a multiple of $p$, and the Schmidt coefficients $\lambda(L,n,k_1,k_2,M_1,M_2)$ take the form
	\begin{equation}
		\lambda(L,n,k_1,k_2,M_1,M_2)=C_{M_1}^{k_1}C_{M_2}^{k_2}\frac{Z_p(n,k_1,k_2)Z_p(L-n,M_1-k_1,M_2-k_2)}{Z_p(L,M_1,M_2)}.
		\label{m1m2p}
	\end{equation}
Here $Z_p(n,k_1,k_2)$ and $Z_p(L-n,M_1-k_1,M_2-k_2)$ take the same form as $Z_p(L,M_1,M_2)$ in Eq.~(\ref{zm1m2p}).
	Therefore, $\lambda(L,n,k_1,k_2,M_1,M_2)$ may be rewritten as
	\begin{equation}
		\lambda(L,n,k_1,k_2,M_1,M_2)=\sqrt{\frac{C_{\gamma n}^{k_1}C_{n-k_1}^{k_2}C_{\gamma(L-n)}^{M_1-k_1}C_{\gamma (L-n)-M_1+k_1}^{M_2-k_2}} {C_{\gamma L}^{M_1}C_{\gamma L-M_1}^{M_2}}}.
		\label{lamm1m2p}
	\end{equation}

In the dimer-trimer regime  $\pi < \theta < 3\pi/2$, the degenerate ground states $|L,M\rangle_p$ admit the exact Schmidt decomposition
\begin{equation}
	|L,M\rangle_p= \sum\limits_{k=0}^{\gamma n}\lambda(L,n,k,M)
	|n,k\rangle_p|L-n,M-k\rangle_p, \label{lmsvd}
\end{equation}
where $n$ is a multiple of $p$, and the Schmidt coefficients $\lambda(L,n,k,M)$ take the form
\begin{equation}
	\lambda(L,n,k,M)=C_{M}^{k}\frac{Z_p(n,k)Z_p(L-n,M-k)}{Z_p(L,M)}.
	\label{Lnmk}
\end{equation}
Here $Z_p(n,k)$ and $Z_p(L-n,M-k)$ take the same form as $Z_p(L,M)$ in Eq.~(\ref{zplmsu2}).
Therefore  $\lambda(L,n,k,M)$ may be rewritten as
\begin{equation}
	\lambda(L,n,k,M)\!=\!\sqrt{\frac{\sum_{\nu=0}^{{\rm min}((1-\gamma) n,k)}\!\sum_{\mu=0}^{{\rm min}((1-\gamma) (L-n),k)}\!2^{n_0+l_0-(1-\gamma) L+2\nu+2\mu}C_{(1-\gamma) n}^\nu{\sum}'_{n_{+},n_{0},\;n_{-}}\!C_{\gamma n}^{n_--\nu}C_{\gamma n-n_{-}+\nu}^{n_0-\gamma n+\nu}C_{(1-\gamma)  (L-n)}^\nu{\sum}'_{l_{+},l_{0},\;l_{-}}\!C_{\gamma  (L-n)}^{l_--\mu}C_{\gamma  (L-n)-l_{-}+\mu}^{l_0-\gamma  (L-n)+\mu}} {\sum_{t=0}^{{\rm min}((1-\gamma) L,M)}2^{N_0-(1-\gamma) L+2t}C_{(1-\gamma) L}^t{\sum}'_{N_{+},N_{0},\;N_{-}}C_{\gamma L}^{N_--t}C_{\gamma L-N_{-}+t}^{N_0-\gamma L+t}}}.
	\label{lamm1m3}
\end{equation}
\end{widetext}

The presence of an exact Schmidt decomposition reflects the self-similarities underlying an abstract fractal in the ground state subspace, which is characterized in terms of the fractal dimension $d_f$,  as already introduced in a field-theoretic approach to the ${\rm SU} (2)$ Heisenberg ferromagnetic states~\cite{doyon}. This is due to the fact that both the block $B$ and the environment $E$, as the two subsystems, share the same type of quantum states as the entire system as a whole. Hence the fractal dimension $d_f$ furnishes a proper description for this type of scale invariant states arising from SSB with type-B GMs.

\section{The entanglement entropy, the number of type-B Goldstone modes and the fractal dimension}~\label{ee}
We are now in a position to evaluate the entanglement entropy for $|L,M_1,M_2\rangle_p$ at the trimer point and for $|L,M\rangle_p$ in the dimer-trimer regime, 
given that  the entanglement entropy for degenerate ground states at the dimer point has been investigated previously in Ref.~\cite{golden}.

\begin{figure}[htp]
	\centering
	\includegraphics[width=0.5\textwidth]{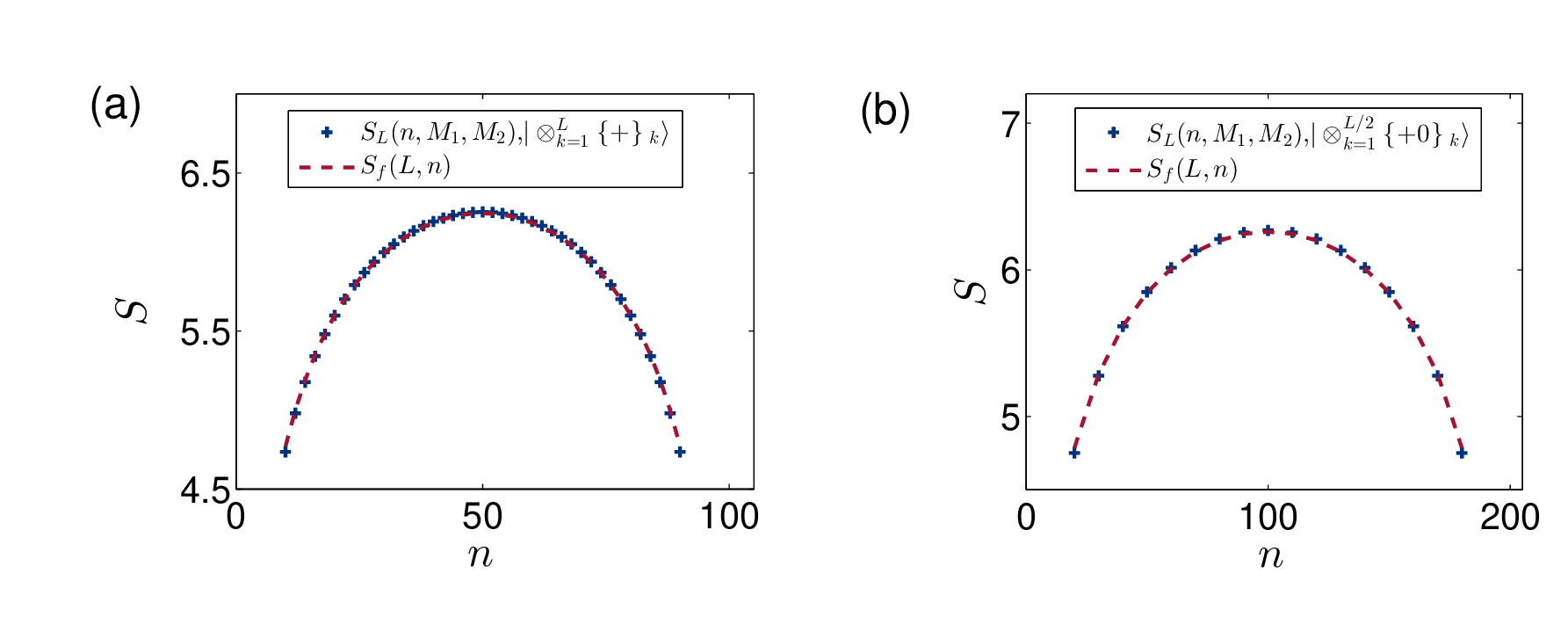}
	\includegraphics[width=0.5\textwidth]{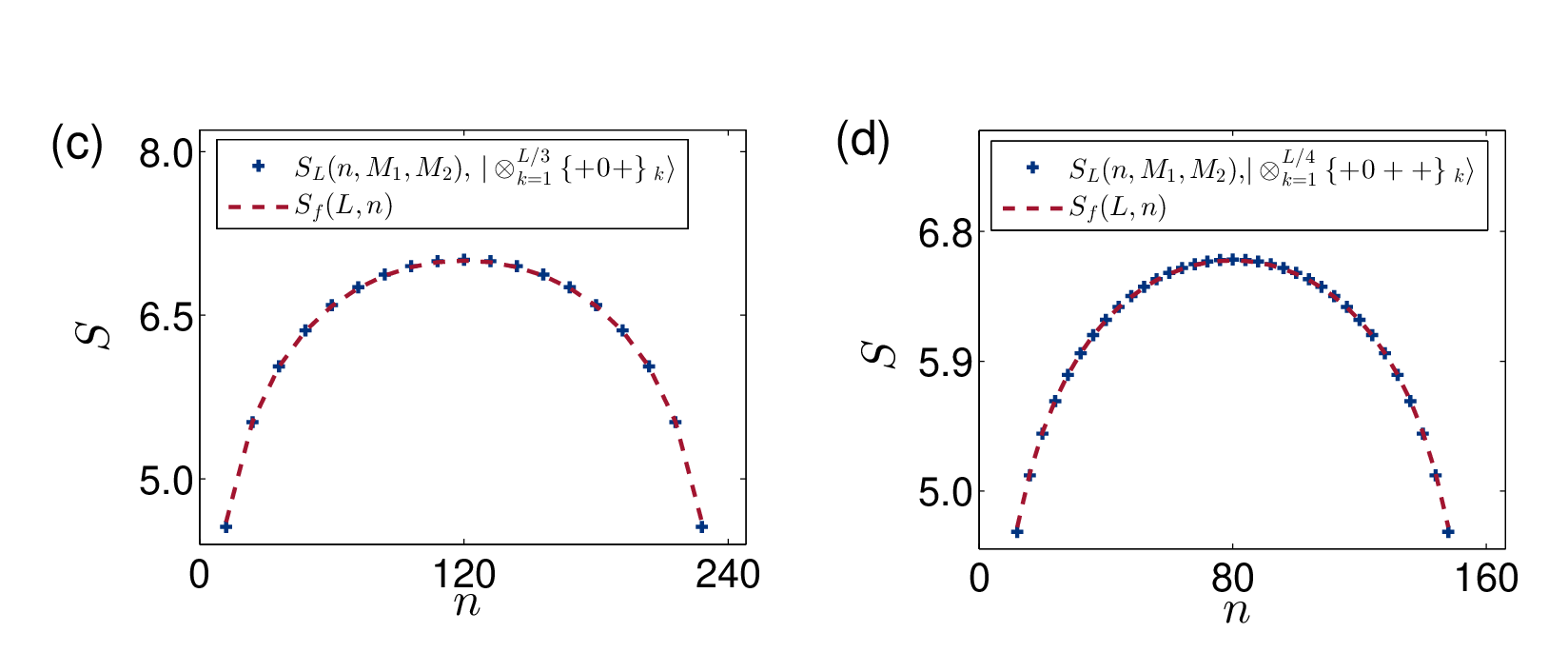}
	\includegraphics[width=0.5\textwidth]{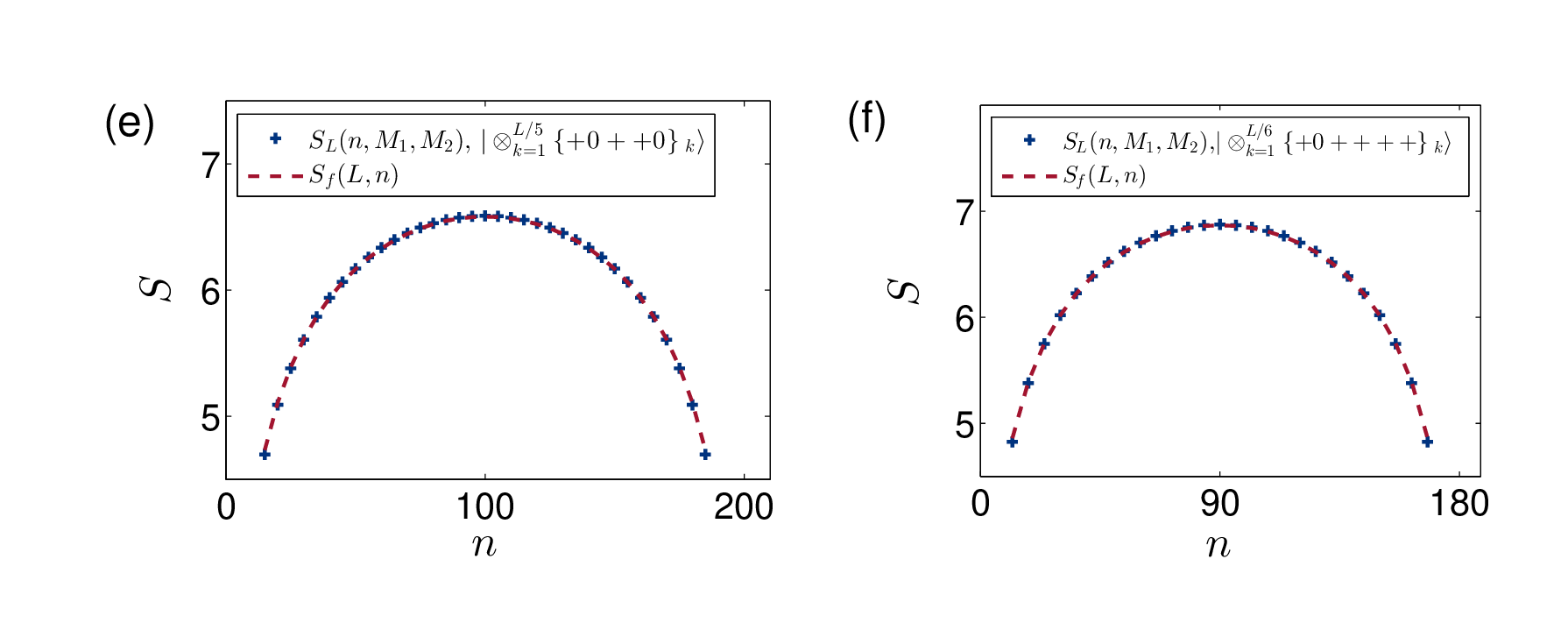}
	\caption{  The entanglement entropy $S_L(n,M_1,M_2)$ versus $n$ at the trimer point for (a) $|L,M_1,M_2\rangle_1$ with fillings $f_1=1/4$ and  $f_2=1/4$, when $L=100$; (b)  $|L,M_1,M_2\rangle_2$ with $f_1=1/5$ and  $f_2=1/5$, when $L=200$;
		(c) $|L,M_1,M_2\rangle_3$ with $f_1=1/4$ and  $f_2=1/4$, when $L=240$; (d) $|L,M_1,M_2\rangle_4$  with $f_1=1/4$ and  $f_2=1/5$, when $L=160$; (e)  $|L,M_1,M_2\rangle_5$ with  $f_1=1/4$ and  $f_2=1/5$, when $L=200$; (f)   $|L,M_1,M_2\rangle_6$ with  $f_1=1/4$ and  $f_2=1/5$, when $L=180$.  The dashed line indicates the entanglement entropy $S_{\!\!f}(L,n)$ versus $n$  evaluated from the universal finite system-size scaling function. The best fitting gives (a)  $S_{\!f0}=1.6002$, (b) $S_{\!f0}=0.6155$, (c)  $S_{\!f0}=1.094$, (d) $S_{\!f0}=1.2775$, (e)  $S_{\!f0}=0.9379$ and (f) $S_{\!f0}=1.3753$.
		Here  (a) $\vert\otimes_{k=1}^{L} \{+\}_{\;k}\rangle$ corresponds to the highest weight state, with period $p=1$ and (b) $\vert\otimes_{k=1}^{L/2} \{+0\}_{\;k}\rangle$, (c) $\vert\otimes_{k=1}^{L/3}  \{+0+\}_{\;k}\rangle$, (d) $\vert\otimes_{k=1}^{L/4}  \{+0++\}_{\;k}\rangle$, (e) $\vert\otimes_{k=1}^{L/5}  \{+0++0\}_{\;k}\rangle$ and (f) $\vert\otimes_{k=1}^{L/6}  \{+0++++\}_{\;k}\rangle$ correspond to generalized highest weight states, with the period $p$ respectively being 2, 3, 4, 5 and 6.
	}
	\label{comparetrimer}
\end{figure}

\begin{figure}[ht]
	\centering
	\includegraphics[width=0.5\textwidth]{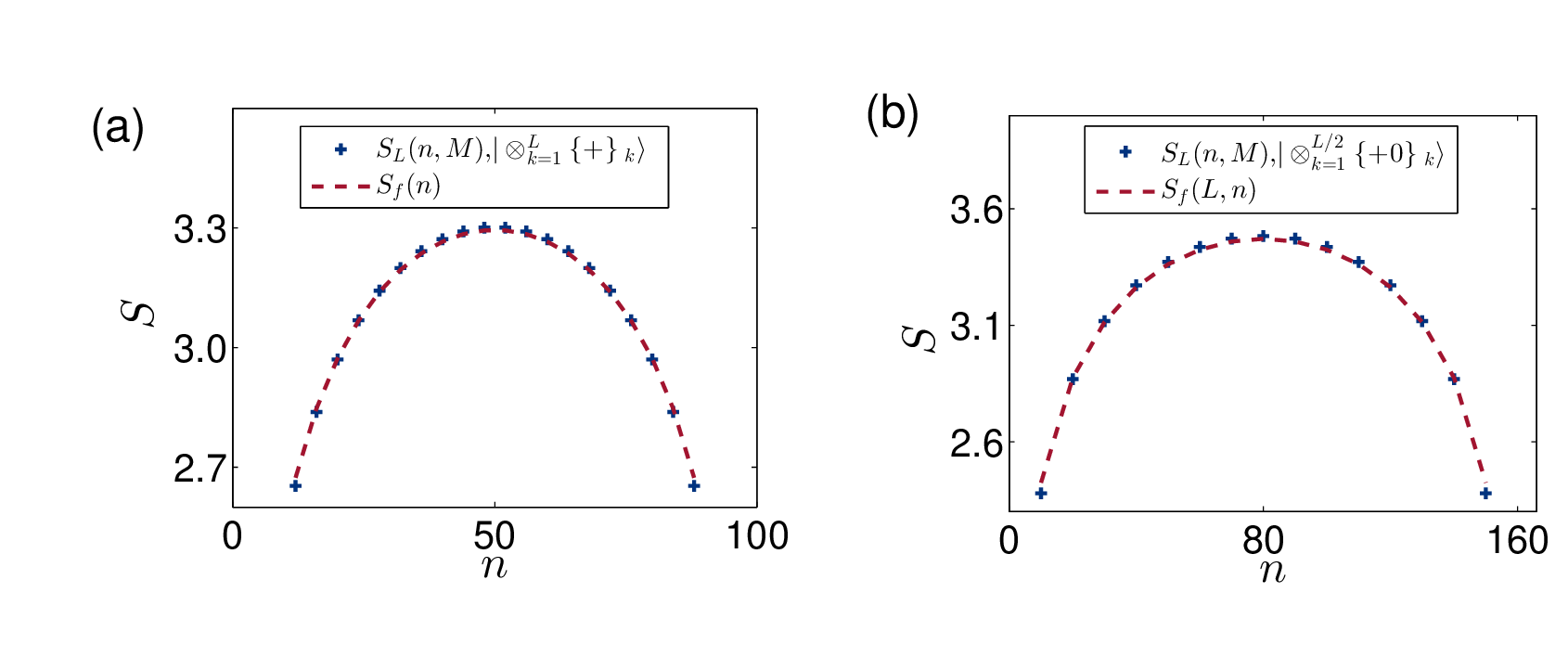}
	\includegraphics[width=0.5\textwidth]{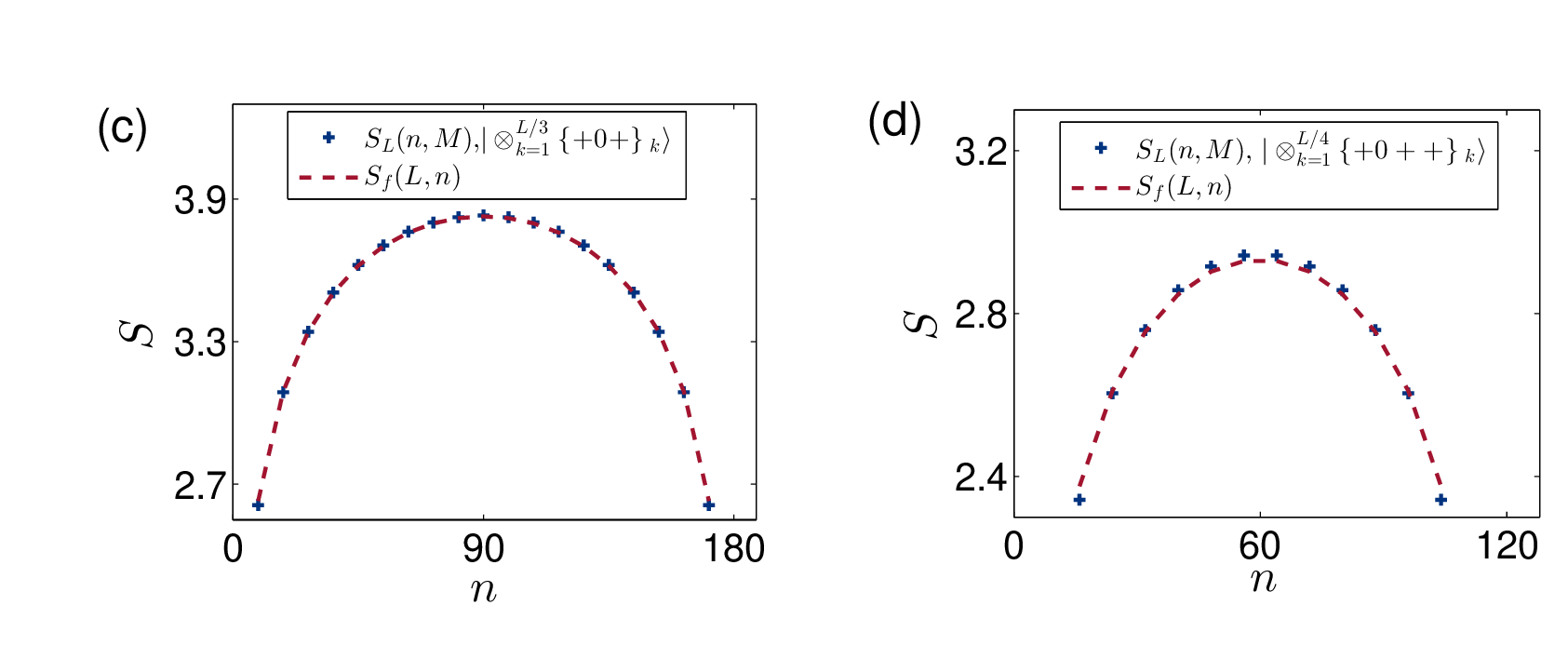}
	\includegraphics[width=0.5\textwidth]{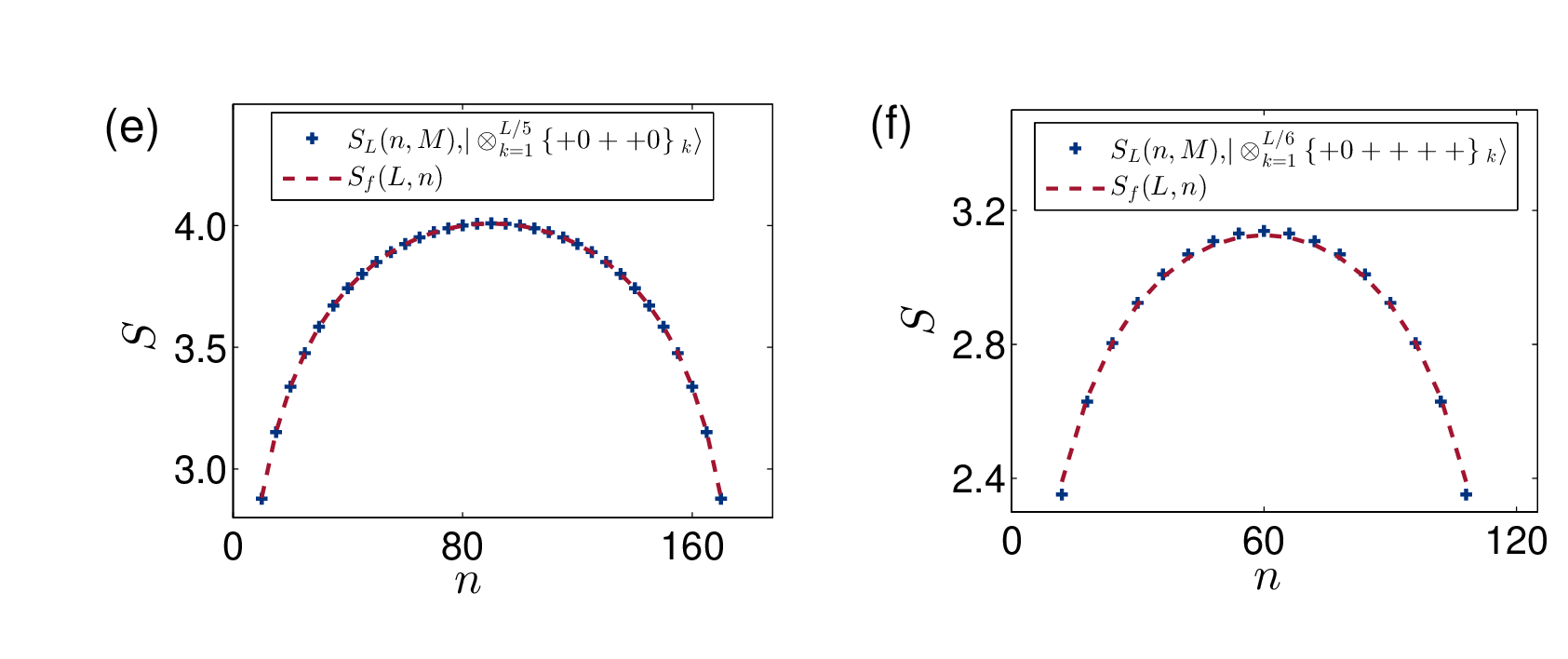}
	\caption{  The entanglement entropy $S_L(n,M)$ versus $n$  in the dimer-trimer regime for (a) $|L,M\rangle_1$ with filling $f=1/4$,  when $L=100$; (b) $|L,M\rangle_2$ with $f=1/5$ when $L=160$; (c) $|L,M\rangle_3$ with $f=1/3$ when $L=180$;  
    (d) $|L,M\rangle_4$ with $f=1/8$ when $L=120$;  
    (e) $|L,M\rangle_5$ with $f=1/2$ when $L=180$; 
    (f) $|L,M\rangle_6$ with $f=1/6$ when $L=120$. The dashed line indicates the entanglement entropy $S_{\!\!f}(L,n)$ versus $n$ evaluated from the universal finite system-size scaling function. The best fitting gives (a)  $S_{\!f0}=0.9733$, (b)  $S_{\!f0}=0.8100$, (c) $S_{\!f0}=1.0811$, (d)  $S_{\!f0}=0.4789$, (e) $S_{\!f0}=1.2629$, and (f)  $S_{\!f0}=0.6735$.
	Here  (a) $\vert\otimes_{k=1}^{L} \{+\}_{\;k}\rangle$ corresponds to the highest weight state, with period $p=1$ and (b) $\vert\otimes_{k=1}^{L/2} \{+0\}_{\;k}\rangle$, (c) $\vert\otimes_{k=1}^{L/3}  \{+0+\}_{\;k}\rangle$, (d) $\vert\otimes_{k=1}^{L/4}  \{+0++\}_{\;k}\rangle$, (e) $\vert\otimes_{k=1}^{L/5}  \{+0++0\}_{\;k}\rangle$ and (f) $\vert\otimes_{k=1}^{L/6}  \{+0++++\}_{\;k}\rangle$ correspond to generalized highest weight states, with the period $p$ respectively being 2, 3, 4, 5 and 6.}
	\label{manypsu2}
\end{figure}

For degenerate ground states $|L,M_1,M_2\rangle_p$, the entanglement entropy $S_{\!L}(n,M_1,M_2)$ follows from
	\begin{align}
		S_{\!L}(n,M_1,M_2)=&-\sum_{k_1,k_2} \Lambda(L,n,k_1,k_2,M_1,M_2) \, \times \nonumber \\
		&\log_{2}\Lambda(L,n,k_1,k_2,M_1,M_2),
		\label{snkm1m2}
	\end{align}
where $\Lambda(L,n,k_1,k_2,M_1,M_2)$ are the eigenvalues of the reduced density matrix $\rho_L(n,M_1,M_2)$: $\Lambda(L,n,k_1,k_2,M_1,M_2)=[\lambda(L,n,k_1,k_2,M_1,M_2)]^2$.
At the trimer point the entanglement entropy $S_{\!L}(n,M_1,M_2)$ may therefore be evaluated as a function of the block size $n$ for fixed $L$, $M_1$ and $M_2$.

For degenerate ground states $|L,M\rangle_p$, the entanglement entropy $S_{\!L}(n,M)$ follows from
\begin{align}
	S_{\!L}(n,M)=-\sum_{k} \Lambda(L,n,k,M)	\log_{2} \Lambda(L,n,k,M),
	\label{snkm1m2}
\end{align}
where $ \Lambda(L,n,k,M)$ are the eigenvalues of the reduced density matrix $\rho_L(n,M)$: $ \Lambda(L,n,k,M)=[\lambda(L,n,k,M)]^2$.
The entanglement entropy $S_{\!L}(n,M)$ in the dimer-trimer regime may therefore be evaluated as a function of the block size $n$ for fixed $L$ and $M$.

In addition, for a scale invariant state arising from SSB with type-B GMs, a universal finite system-size scaling function is exposed~\cite{finitesize}. Hence the entanglement entropy $S_{\!\!f}(L,n)$ scales as
\begin{equation}
	S_{\!\!f}(L,n)=\frac{N_B}{2} \log_2\frac{n(L-n)}{L} +S_{\!\!f0}.
	\label{slnf}
\end{equation}
Here the subscript $f$  refers to a set of fillings  $f_\alpha$ ($\alpha =1, \ldots,r$), defined as $f_\alpha = M_\alpha/L$, where $r$ denotes the rank of the (semisimple) symmetry group, and   $S_{\!\!f0}$ is an additive non-universal constant~\cite{finitesize,SU4}. For ${\rm SU}(2)$, the rank $r=1$ and for ${\rm SU}(3)$, the rank $r=2$.

We remark that the finite system-size scaling relation for the entanglement entropy $S_{\!\!f}(L,n)$, as presented in Eq.~(\ref{slnf}), becomes a logarithmic scaling relation for  $S_{\!\!f}(n)$ with block size $n$  in the thermodynamic limit $L \rightarrow \infty$, consistent with a generic but heuristic argument given in Refs.~\cite{FMGM,finitesize}. Here $S_{\!\!f}(n)$ is defined as $S_{\!\!f}(L,n)$ when the system size $L$ tends to infinity. That is, we have
\begin{equation}
S_{\!\!f}(n)=\frac {N_B}{2}\log_2n+S_{\!\!f0}.
\end{equation}
Combined with the field-theoretic prediction
\begin{equation}
S_{\!\!f}(n)=\frac{d_f}{2}\log_2n+S_{\!\!f0},
\end{equation}
made by Castro-Alvaredo and Doyon~\cite{doyon} for the ${\rm SU}(2)$ ferromagnetic states,
the fractal dimension $d_f$ is identical to the number of type-B GMs $N_B$, namely
\begin{equation}
d_f=N_B.
\end{equation}
The fractal dimension $d_f$ is thus an integer for linearly independent degenerate ground states generated from the repeated action of the lowering operators on the highest weight state and generalized highest weight states, as extracted from a finite system-size scaling analysis. We stress that the fractal dimension $d_f$ characterizes an abstract fractal underlying the ground state subspace in the Hilbert space, which in turn reflects the self-similarities behind degenerate ground states. However, it appears to be necessary to look for a physical observable that reflects the self-similarities. As it turns out, the residual entropy is a proper choice for this purpose. Indeed, the self-similar geometric objects characterizing asymptotic behaviors of the ground state degeneracies for large $L$ are logarithmic spirals, with the golden spiral as a special case at the dimer point \cite{golden}.

In Fig.~\ref{comparetrimer}, we plot the entanglement entropy $S_{\!L}(n,M_1,M_2)$ versus $n$ at the trimer point for the indicated states and fillings.
The entanglement entropy $S_{\!\!f}(L,n)$ versus $n$, evaluated from the universal finite system-size scaling function, is indicated as a dashed line. Our numerical data for $S_{\!L}(n,M_1,M_2)$ fall onto the curve $S_{\!\!f}(L,n)$, with the relative errors less than $1.5\%$. Here and hereafter, we have regarded $S_{\!\!f}(L,n)$ as a function of $n$ for fixed $L$ and $f$.

Fig.~\ref{manypsu2} shows similar plots of the entanglement entropy $S_{\!L}(n,M)$ versus $n$ in the dimer-trimer regime. The entanglement entropy $S_{\!\!f}(L,n)$ versus $n$, evaluated from the universal finite system-size scaling function, is indicated as a dashed line. Our numerical data for $S_{\!L}(n,M)$ fall onto the curve $S_{\!\!f}(L,n)$, with the relative errors less than $2\%$.

\section{Summary}\label{sum}
We have systematically investigated SSB with type-B GMs in the macroscopically degenerate phase  of the quantum spin-1 model (\ref{model}) with competing dimer and trimer interactions, including the QPT points located at the two endpoints. It is found that the model involves three distinct SSB patterns. The first occurs at the dimer point ($\theta = \pi$), with the pattern from the staggered symmetry group ${\rm SU}(3)$  to ${\rm U}(1)\times{\rm U}(1)$, so the number of type-B GMs is two. The second  occurs at the trimer point ($\theta = 3\,\pi/2$), with the pattern from the  uniform symmetry group ${\rm SU}(3)$ to ${\rm U}(1)\times{\rm U}(1)$, so the number of type-B GMs is two. The third occurs in the dimer-trimer regime ($\pi < \theta < 3\, \pi/2$), with the pattern from the  uniform symmetry group ${\rm SU}(2)$ to ${\rm U}(1)$, so the number of type-B GMs is one. The ground state degeneracies arising from the three patterns are exponential with the system size, which may be recognized as sequences of integers relevant to self-similar geometric objects. These are logarithmic spirals, with the golden spiral as a special case at the dimer point~\cite{golden}.

The presence of an emergent symmetry operation tailored to a specific degenerate ground state is responsible for the fact that the ground state degeneracies $\Omega_L^{\rm OBC/PBC}$ are exponential as a function of the system size $L$.
 As a consequence, the residual entropy $S_r$ is non-zero. Physically, the residual entropy measures the disorder present in a unit cell of highly degenerate ground state generated from a generalized highest weight state.

Meanwhile, an exact Schmidt decomposition for highly degenerate ground states exposes the self-similarities, an essential feature underlying an abstract fractal in the ground state subspace described by the fractal dimension. A universal finite system-size scaling analysis of the entanglement entropy has been performed for the highly degenerate ground states, in order to extract the fractal dimension, which is identical to the number of type-B GMs.

As a result of the three SSB patterns, the dimer and trimer models under consideration accommodate scale invariant, but not conformally invariant ground states that represent an exotic quantum state of matter, featuring that the translation-invariant ground states coexist with dimerized ground states, trimerized ground states, tetramerized  ground states and so on in the dimer-trimer regime and at the trimer point. 
More concisely, the translation-invariant ground states coexist with $p$-merized states, where  integer $p \ge 2$. 
In contrast, at the dimer point, the translation-invariant ground states coexist with $p$-merized states where $p$ is even.

\section{Acknowledgements}
We thank  John Fjaerestad for enlightening discussions about the counting rule for Goldstone modes and Hosho Katsura for bringing our attention to Ref.~\cite{dtmodel2}.
I.P.M. acknowledges funding from the National Science and Technology Council (NSTC) Grant No. 122-2811-M-007-044.

%%%%%%%%%%%%%%%%%%%%%%%%%%%%%%%%%%%%%%%%%%%%%%%%%Appendix%%%%%%%%%%%%%%%%%%%%%%%5
\newpage
\onecolumngrid
\newpage
\section*{Appendix}
\twocolumngrid
\setcounter{equation}{0}
\setcounter{figure}{0}
\renewcommand{\theequation}{S\arabic{equation}}
\renewcommand{\thefigure}{S\arabic{figure}}
\renewcommand{\bibnumfmt}[1]{[S#1]}
\renewcommand{\citenumfont}[1]{S#1}

\subsection{The staggered ${\rm SU(3)}$ symmetry group at the dimer point, the uniform ${\rm SU(3)}$  symmetry group at the trimer point and the uniform ${\rm SU(2)}$ symmetry group  in the dimer-trimer regime}\label{symmetry}

Here we summarize the generators and their commutation relations for the symmetry groups as the parameter $\theta$ varies in the macroscopically degenerate phase, including the two ends representing the two QPT points. That is, the staggered ${\rm SU(3)}$ symmetry group emerges at the dimer point $\theta=\pi$, the uniform ${\rm SU(3)}$ symmetry group  emerges at the trimer point $\theta=3\pi/2$ and the uniform ${\rm SU(2)}$ symmetry group emerges in the dimer-trimer regime ($\pi <\theta < 3\pi/2$).

(a) Model (1) at the dimer point ($\theta=\pi$) possesses the staggered ${\rm SU(3)}$ symmetry group, with its generators being the eight traceless matrices $J_\delta = \sum_j J_{\delta,j}$ ($\delta = 1, \ldots, 8$),  where
$J_{1,j}=S^x_j/2$, $J_{2,j}=S^y_j/2$, $J_{3,j}=S^z_j/2$, $J_{4,j}=(-1)^j(I-3/2{S^z_j}^2)$, $J_{5,j}=(-1)^j[{(S^x_j)}^2-{(S^y_j)}^2]/2$, $J_{6,j}=(-1)^j(S^x_jS^y_j+S^y_jS^x_j)/2$, $J_{7,j}=(-1)^j(S^y_jS^z_j+S^z_jS^y_j)/2$, and $J_{8,j}=(-1)^j(S^z_jS^x_j+S^x_jS^z_j)/2$~\cite{LLspin1, golden}. Here $I$ is the $3 \times 3$ identity matrix.

The rank $r$ is two. Accordingly, there are two Cartan generators $H_1$ and $H_2$, which are traceless and diagonal:
$H_1=\sum_jH_{1,j}$ and $H_2=\sum_jH_{2,j}$,
where $H_{1,j}=S^z_j/2+(-1)^j(I-3/2{S^z_j}^2)$, and $H_{2,j}=2 S^z_j$. For each of $H_\alpha$ ($\alpha =  1$ and 2),  there exists a conjugate pair of a raising operator $E_{\alpha}=\sum_jE_{\alpha,j}$ and a lowering operator $F_{\alpha}=\sum_jF_{\alpha,j}$,
with $E_{1,j} = \sqrt{2}/{4}[S^x_j-(-1)^j(S^z_jS^x_j+S^x_jS^z_j)
+iS^y_j- i (-1)^j(S^y_jS^z_j+S^z_jS^y_j)]$, $E_{2,j}= (-1)^{j+1}[{(S^x_j)}^2-{(S^y_j)}^2+i(S^x_jS^y_j+S^y_jS^x_j)]/2$,
	$F_{1,j}=\sqrt{2}/{4}[S^x_j-(-1)^j(S^z_jS^x_j+S^x_jS^z_j)
	-iS^y_j+i(-1)^j(S^y_jS^z_j+S^z_jS^y_j)]$ and $F_{2,j}=-(-1)^j[{(S^x_j)}^2-{(S^y_j)}^2-i(S^x_jS^y_j+S^y_jS^x_j)]/2$.
In addition, there are two extra generators: a raising operator $E_3=\sum_jE_{3,j}$ and a lowering operator $F_3=\sum_jF_{3,j}$, with
   $E_{3,j}=\sqrt{2}/{4}[S^x_j+(-1)^j(S^z_jS^x_j+S^x_jS^z_j)
   +iS^y_j+i(-1)^j(S^y_jS^z_j+S^z_jS^y_j)]$, $F_{3,j}= \sqrt{2}/{4}[S^x_j+(-1)^j(S^z_jS^x_j+S^x_jS^z_j)
   -iS^y_j-i(-1)^j(S^y_jS^z_j+S^z_jS^y_j))]$.
They satisfy the commutation relations 
$[H_1,E_1]=2E_1$, $[H_1,F_1]=-2F_1$, $[E_1,F_1]=H_1$,
$[H_2,E_2]=2E_2$, $[H_2,F_2]=-2F_2$, $[E_2,F_2]=H_2$,
$[H_2-H_1,E_3]=2E_3$, $[H_2-H_1,F_3]=-2F_3$,
$[E_3,F_3]=H_2-H_1$, $[H_1,E_2]=E_2$,
$[H_1,E_3]=-E_3$, $[H_2,E_1]=E_1$, $[H_2,E_3]=E_3$,
$[H_1,F_2]=-F_2$, $[H_1,F_3]=F_3$, $[H_2,F_1]=-F_1$,
$[H_2,F_3]=-F_3$, $[F_1,F_2]=0$, $[F_2,F_3]=0$,
$[E_1,E_2]=0$ and $[E_1,E_3]=0$.

Denoting $|+\rangle$ as the eigenvectors of the spin operator $S^z_j$, with eigenvalue $1$, we have
$E_1 \vert\otimes_{k=1}^{L} \{+\}_{\;k}\rangle=0$,  $E_2\vert\otimes_{k=1}^{L} \{+\}_{\;k}\rangle=0$ and $E_3\vert\otimes_{k=1}^{L} \{+\}_{\;k}\rangle=0$,  together with $H_1\vert\otimes_{k=1}^{L} \{+\}_{\;k}\rangle=H_2\vert\otimes_{k=1}^{L} \{+\}_{\;k}\rangle=\vert\otimes_{k=1}^{L} \{+\}_{\;k}\rangle$.
Therefore, $\vert\otimes_{k=1}^{L} \{+\}_{\;k}\rangle$ is by definition the highest weight state.
As argued in Ref.~\cite{golden}, there are six broken generators, which only yield two type-B GMs.

(2) Model  (1) at the trimer point ($\theta=3\pi/2$) possesses the uniform ${\rm SU(3)}$ symmetry group. The rank $r$ is two.
Accordingly, there are two Cartan generators $H_1$ and $H_2$,  which may be chosen as $H_{1,j}=3/2(S^z_j)^2+S^z_j/2-I$ and $H_{2,j}=S^z_j$.
For each of the $H_1$ and $H_2$, a lowering operator and a raising operator may be chosen as $F_1=\sum_jF_{1,j}$,
$F_2=\sum_jF_{2,j}$, $E_1=\sum_jE_{1,j}$, $E_2=\sum_jE_{2,j}$, with $F_{1,j}=\sqrt{2}/2(S_j^x+S_j^xS_j^z-iS_j^zS_j^y)$, $F_{2,j}=[(S^x_j)^2-(S^y_j)^2 -i(S^x_jS^y_j+S^y_jS^x_j)]/2$, $E_{1,j}=\sqrt{2}/2(S_j^x+S_j^zS_j^x+iS_j^yS_j^z)$ and $E_{2,j}=[(S^x_j)^2-(S^y_j)^2 +i(S^x_jS^y_j+S^y_jS^x_j)]/2$.
In addition, there are two extra generators: a raising operator $E_3=\sum_jE_{3,j}$ and a lowering operator $F_3=\sum_jF_{3,j}$, with
$E_{3,j}=\sqrt{2}/{2}[S^x_j-S^z_jS^x_j-i S^z_jS^y_j]$, $F_{3,j}= \sqrt{2}/{2}[S^x_j-S^x_jS^z_j+i S^y_jS^z_j]$.
They satisfy the commutation relations $[H_1,E_1]=2E_1$, $[H_1,F_1]=-2F_1$, $[E_1,F_1]=H_1$,
$[H_2,E_2]=2E_2$, $[H_2,F_2]=-2F_2$, $[E_2,F_2]=H_2$,
$[H_2-H_1,E_3]=2E_3$, $[H_2-H_1,F_3]=-2F_3$,
$[E_3,F_3]=H_2-H_1$, $[H_1,E_2]=E_2$,
$[H_1,E_3]=-E_3$, $[H_2,E_1]=E_1$, $[H_2,E_3]=E_3$,
$[H_1,F_2]=-F_2$, $[H_1,F_3]=F_3$, $[H_2,F_1]=-F_1$,
$[H_2,F_3]=-F_3$, $[F_1,F_2]=0$, $[F_2,F_3]=0$,
$[E_1,E_2]=0$ and $[E_1,E_3]=0$.

Note that $E_1 \vert\otimes_{k=1}^{L} \{+\}_{\;k}\rangle=0$, $E_2\vert\otimes_{k=1}^{L} \{+\}_{\;k}\rangle=0$ and $E_3\vert\otimes_{k=1}^{L} \{+\}_{\;k}\rangle=0$,  together with  $H_1\vert\otimes_{k=1}^{L} \{+\}_{\;k}\rangle=H_2\vert\otimes_{k=1}^{L} \{+\}_{\;k}\rangle=\vert\otimes_{k=1}^{L} \{+\}_{\;k}\rangle$.
Therefore  $\vert\otimes_{k=1}^{L} \{+\}_{\;k}\rangle$ is by definition the highest weight state.
The interpolating fields are $E_{1,j}$ and $F_{1,j}$ for the generators $F_{1}$ and $E_{1}$,  $E_{2,j}$ and $F_{2,j}$ for the generators $F_2$ and $E_2$,  and $E_{3,j}$ and $F_{3,j}$ for the generators $F_{3}$ and $E_{3}$, respectively.
Thus $\langle H_{1,j}\rangle$ and $\langle H_{2,j}\rangle$ are the local order parameters, given  $\langle[E_{1,j},F_1]\rangle=\langle[E_1,F_{1,j}]\rangle=\langle H_{1,j}\rangle\neq0$, $\langle[E_{2,j},F_2]\rangle=\langle[E_2,F_{2,j}]\rangle=\langle H_{2,j}\rangle\neq0$,
and $\langle[E_{3,j},F_3]\rangle=\langle[E_3,F_{3,j}]\rangle=\langle H_{2,j}-H_{1,j}\rangle=0$.
Hence the four symmetry generators $E_1$, $E_2$,  $F_1$ and $F_2$ are spontaneously broken.
According to the counting rule~\cite{watanabe}, two type-B GMs emerge.

(3) Model  (1) in the dimer-trimer regime ($\pi <\theta < 3\pi/2$)  possesses the uniform ${\rm SU(2)}$ symmetry group. The rank is one. Accordingly, there is one Cartan generator $S^z$, defined as $S^z=\sum_jS^z_j$.  For $S^z$ there exists a conjugate pair of a lowering operator $S^-$ and a raising operator $S^+$,
defined as $S^+=\sum_jS^+_j$ and $S^-=\sum_jS^-_j$. They satisfy the commutation relations $[S^z,S^+]=S^+$, $[S^+,S^-]=S^z$ and $[S^-,S^z]=S^-$, where $S^+_j$ and $S^-_j$ are defined as $ S^{\pm}_j=(S^x_j\pm iS^y_j)/\sqrt{2}$, with $S^x_j$, $S^y_j$, and $S^z_j$  the spin-1 operators at the $j$-th site.
The action of   $S^z_j$  and $S^+_j$ on $|+\rangle_j$ takes the form  $S^z_j|+\rangle_j=|+\rangle_j$ and $S^+_j|+\rangle_j=0$, respectively.
Therefore $\vert\otimes_{k=1}^{L} \{+\}_{\;k}\rangle$ is by definition  the highest weight state.
The interpolating fields are $S^-_j$ and $S^+_j$ for the generators $S^+$ and $S^-$,  respectively.
Hence $\langle S^z_j\rangle$ is the local order parameter, given $\langle[S^-_j,S^+]\rangle=\langle[S^+_j,S^-]\rangle=\langle S^z_j\rangle\neq0$.
Thus the two symmetry generators $S^x$ and $S^y$ are spontaneously broken.
According to the counting rule~\cite{watanabe}, one type-B GM emerges.

\begin{widetext}
\subsection{An explicit expression for the commutator $V$ between the Hamiltonian and a local unitary operation}

The degenerate ground state $|\psi_0\rangle=S^-\vert\otimes_{k=1}^{L} \{+\}_{\;k}\rangle$ is an eigenvector of $S^z$ with the eigenvalue $L-1$.   We consider a set of local unitary operations $g_j$, defined as $\exp (i \pi \Sigma_j)$, where $\Sigma_j =S_j^z$, with $j=1,\ldots, L$ under PBCs and $j=2,\ldots, L-1$ under OBCs. Then at the dimer point ($\theta=\pi/2$),
the commutators between the Hamiltonian $H$ and the local unitary operations $g_j$, denoted as $V_j^D$, take the form
	\begin{align*}
		V_j^D= &- (e_{  2   3})_j (e_{  2   1})_{j+1}
		+ (e_{  3   2})_j (e_{  1   2})_{j+1}
		+ (e_{  1   2})_j (e_{  3   2})_{j+1}
		- (e_{  2   1})_j (e_{  2   3})_{j+1}   \\
&		- (e_{  2   3})_{j-1} (e_{  2   1})_j
		+ (e_{  3   2})_{j-1} (e_{  1   2})_j
		+ (e_{  1   2})_{j-1} (e_{  3   2})_j
		- (e_{  2   1})_{j-1} (e_{  2   3})_j .  \\
	\end{align*}
Here we have introduced the basis matrices $e_{uv}$, defined as $e_{uv}=|u\rangle\langle v|$, with $|u\rangle$ and $|v\rangle$ being the $u$-th and $v$-th states in an orthonormal basis, where $u,v=1,2$ and $3$. The latter is identified with the basis consisting of three basis states, with the eigenvalues of spin-1 $S_j^z$ being 1, 0 and $-1$. Hence we have
\begin{align*}
	e_{11}=&\frac{S^z+(S^z)^2}{2}, \quad e_{12}=\frac{\sqrt{2}(S^zS^x+(S^z)^2S^x)}{2},  \quad e_{13}=\frac{(S^z+(S^z)^2)((S^x)^2+iS^xS^y)}{2},\nonumber \\
	e_{21}=&\frac{\sqrt{2}[S^x(S^z+(S^z)^2)]}{2}, \quad e_{22}=S^x(S^z)^2S^x, \quad 
	e_{23}=\frac{\sqrt{2}[(S^x+iS^y)(S^z)^2]}{2}, \nonumber \\
	e_{31}=&\frac{((S^x)^2-iS^yS^x)(S^z+(S^z)^2)}{2},  \quad e_{32}=\frac{\sqrt{2}[(S^z)^2(S^x-iS^y)]}{2}, \quad
	e_{33}=\frac{(S^z)^2-S^z}{2}. \nonumber \\
\end{align*}

At the trimer point ($\theta=3\pi/2$), $g_j$ take the same form as that at the dimer point ($\theta=\pi/2$), with $j=1,\ldots, L$ under PBCs  and $j=3,\ldots, L-2$  under OBCs, the commutators between the Hamiltonian $H$ and the local unitary operations $g_j$, denoted as $V_j^T$, take the form
\begin{align*}	
	V_j^T=&- (e_{  3   3})_j (e_{  1   2})_{j+1} (e_{  2   1})_{j+2}
		+ (e_{  1   3})_j (e_{  3   2})_{j+1} (e_{  2   1})_{j+2}
		+ (e_{  3   2})_j (e_{  1   3})_{j+1} (e_{  2   1})_{j+2}
		- (e_{  1   2})_j (e_{  3   3})_{j+1} (e_{  2   1})_{j+2}  \\
       &+ (e_{  3   3})_j (e_{  2   1})_{j+1} (e_{  1   2})_{j+2}
		- (e_{  2   3})_j (e_{  3   1})_{j+1} (e_{  1   2})_{j+2}
		+ (e_{  2   3})_j (e_{  1   1})_{j+1} (e_{  3   2})_{j+2}
		- (e_{  1   3})_j (e_{  2   1})_{j+1} (e_{  3   2})_{j+2}   \\
	   &- (e_{  3   1})_j (e_{  2   3})_{j+1} (e_{  1   2})_{j+2}
		+ (e_{  2   1})_j (e_{  3   3})_{j+1} (e_{  1   2})_{j+2}
		- (e_{  2   1})_j (e_{  1   3})_{j+1} (e_{  3   2})_{j+2}
		+ (e_{  1   1})_j (e_{  2   3})_{j+1} (e_{  3   2})_{j+2} \\
	&   - (e_{  3   2})_j (e_{  1   1})_{j+1} (e_{  2   3})_{j+2}
		+ (e_{  1   2})_j (e_{  3   1})_{j+1} (e_{  2   3})_{j+2}
		+ (e_{  3   1})_j (e_{  1   2})_{j+1} (e_{  2   3})_{j+2}
		- (e_{  1   1})_j (e_{  3   2})_{j+1} (e_{  2   3})_{j+2} \\
    &   + (e_{  2   3})_{j-1} (e_{  3   2})_{j} (e_{  1   1})_{j+1}
		+ (e_{  3   3})_{j-1} (e_{  1   2})_{j} (e_{  2   1})_{j+1}
		- (e_{  1   3})_{j-1} (e_{  3   2})_{j} (e_{  2   1})_{j+1}
		- (e_{  2   3})_{j-1} (e_{  1   2})_{j} (e_{  3   1})_{j+1}  \\
	&	- (e_{  3   2})_{j-1} (e_{  2   3})_{j} (e_{  1   1})_{j+1}
		+ (e_{  1   2})_{j-1} (e_{  2   3})_{j} (e_{  3   1})_{j+1}
		- (e_{  3   3})_{j-1} (e_{  2   1})_{j} (e_{  1   2})_{j+1}
		+ (e_{  1   3})_{j-1} (e_{  2   1})_{j} (e_{  3   2})_{j+1} \\
   &	+ (e_{  3   1})_{j-1} (e_{  2   3})_{j} (e_{  1   2})_{j+1}
		- (e_{  1   1})_{j-1} (e_{  2   3})_{j} (e_{  3   2})_{j+1}
		+ (e_{  3   2})_{j-1} (e_{  2   1})_{j} (e_{  1   3})_{j+1}
		- (e_{  1   2})_{j-1} (e_{  2   1})_{j} (e_{  3   3})_{j+1} \\
   &    - (e_{  2   1})_{j-1} (e_{  3   2})_{j} (e_{  1   3})_{j+1}
		- (e_{  3   1})_{j-1} (e_{  1   2})_{j} (e_{  2   3})_{j+1}
		+ (e_{  1   1})_{j-1} (e_{  3   2})_{j} (e_{  2   3})_{j+1}
		+ (e_{  2   1})_{j-1} (e_{  1   2})_{j} (e_{  3   3})_{j+1}  \\
  &	    - (e_{  2   3})_{j-2} (e_{  3   2})_{j-1} (e_{  1   1})_j
		+ (e_{  2   3})_{j-2} (e_{  1   2})_{j-1} (e_{  3   1})_j
		+ (e_{  3   2})_{j-2} (e_{  2   3})_{j-1} (e_{  1   1})_j
		- (e_{  3   2})_{j-2} (e_{  1   3})_{j-1} (e_{  2   1})_j  \\
	&   + (e_{  1   2})_{j-2} (e_{  3   3})_{j-1} (e_{  2   1})_j
		- (e_{  1   2})_{j-2} (e_{  2   3})_{j-1} (e_{  3   1})_j
		+ (e_{  2   3})_{j-2} (e_{  3   1})_{j-1} (e_{  1   2})_j
		- (e_{  2   3})_{j-2} (e_{  1   1})_{j-1} (e_{  3   2})_j  \\
	&   - (e_{  2   1})_{j-2} (e_{  3   3})_{j-1} (e_{  1   2})_j
		+ (e_{  2   1})_{j-2} (e_{  1   3})_{j-1} (e_{  3   2})_j
		- (e_{  3   2})_{j-2} (e_{  2   1})_{j-1} (e_{  1   3})_j
		+ (e_{  3   2})_{j-2} (e_{  1   1})_{j-1} (e_{  2   3})_j \\
	& 	- (e_{  1   2})_{j-2} (e_{  3   1})_{j-1} (e_{  2   3})_j
		+ (e_{  1   2})_{j-2} (e_{  2   1})_{j-1} (e_{  3   3})_j
		+ (e_{  2   1})_{j-2} (e_{  3   2})_{j-1} (e_{  1   3})_j
		- (e_{  2   1})_{j-2} (e_{  1   2})_{j-1} (e_{  3   3})_j .
	\end{align*}
%\end{widetext}
Note that under PBCs we denote $(e_{uv})_{-1}\equiv (e_{uv})_{L}$, $(e_{uv})_{-2}\equiv (e_{uv})_{L-1}$, $(e_{uv})_{L+1}\equiv (e_{uv})_{1}$ and $(e_{uv})_{L+2}\equiv (e_{uv})_{2}$ for brevity.

In the MD regime ($\pi <\theta < 3\pi/2$), $g_j$ are identical to that at the trimer point ($\theta=3\pi/2$). The commutators between the Hamiltonian $H$ and the local unitary operations $g_j$, denoted as $V_j^{\rm MD}$, take the form
	\begin{align*}
	V_j^{\rm MD}=-2\cos\theta V_j^D-2\sin\theta V_j^T.
	\end{align*}
We remark that under OBCs $V_j^D$ and  $V_j^T$ take the same form as above, except for $j=1,2, L-1$ and $L$. For these values we have
	
	\begin{align*}
	&V_1^D=- (e_{  2   3})_1 (e_{  2   1})_{2}
		+ (e_{  3   2})_1 (e_{  1   2})_{2}
		+ (e_{  1   2})_1 (e_{  3   2})_{2}
		- (e_{  2   1})_1 (e_{  2   3})_{2},  
   \end{align*}
   \begin{align*}
	&V_1^T=- (e_{  3   3})_1 (e_{  1   2})_2 (e_{  2   1})_3
	+ (e_{  1   3})_1 (e_{  3   2})_2 (e_{  2   1})_3
	+ (e_{  3   2})_1 (e_{  1   3})_2 (e_{  2   1})_3
	- (e_{  1   2})_1 (e_{  3   3})_2 (e_{  2   1})_3  \\
	&+ (e_{  3   3})_1 (e_{  2   1})_2 (e_{  1   2})_3
	- (e_{  2   3})_1 (e_{  3   1})_2 (e_{  1   2})_3
	+ (e_{  2   3})_1 (e_{  1   1})_2 (e_{  3   2})_3
	- (e_{  1   3})_1 (e_{  2   1})_2 (e_{  3   2})_3   \\
	&- (e_{  3   1})_1 (e_{  2   3})_2 (e_{  1   2})_3
	+ (e_{  2   1})_1 (e_{  3   3})_2 (e_{  1   2})_3
	- (e_{  2   1})_1 (e_{  1   3})_2 (e_{  3   2})_3
	+ (e_{  1   1})_1 (e_{  2   3})_2 (e_{  3   2})_3 \\
	&   - (e_{  3   2})_1 (e_{  1   1})_2 (e_{  2   3})_3
	+ (e_{  1   2})_1 (e_{  3   1})_2 (e_{  2   3})_3
	+ (e_{  3   1})_1 (e_{  1   2})_2 (e_{  2   3})_3
	- (e_{  1   1})_1 (e_{  3   2})_2 (e_{  2   3})_3 , 
	\end{align*}
\begin{align*}
	V_2^D=&- (e_{  2   3})_2 (e_{  2   1})_3
	+ (e_{  3   2})_2 (e_{  1   2})_3
	+ (e_{  1   2})_2 (e_{  3   2})_3
	- (e_{  2   1})_2 (e_{  2   3})_3   \\
	&		- (e_{  2   3})_1 (e_{  2   1})_2
	+ (e_{  3   2})_1 (e_{  1   2})_2
	+ (e_{  1   2})_1 (e_{  3   2})_2
	- (e_{  2   1})_1 (e_{  2   3})_2,
\end{align*}
\begin{align*}
	V_2^T=&- (e_{  3   3})_2 (e_{  1   2})_3 (e_{  2   1})_4
	+ (e_{  1   3})_2 (e_{  3   2})_3 (e_{  2   1})_4
	+ (e_{  3   2})_2 (e_{  1   3})_3 (e_{  2   1})_4
	- (e_{  1   2})_2 (e_{  3   3})_3 (e_{  2   1})_4  \\
	&+ (e_{  3   3})_2 (e_{  2   1})_3 (e_{  1   2})_4
	- (e_{  2   3})_2 (e_{  3   1})_3 (e_{  1   2})_4
	+ (e_{  2   3})_2 (e_{  1   1})_3 (e_{  3   2})_4
	- (e_{  1   3})_2 (e_{  2   1})_3 (e_{  3   2})_4   \\
	&- (e_{  3   1})_2 (e_{  2   3})_3 (e_{  1   2})_4
	+ (e_{  2   1})_2 (e_{  3   3})_3 (e_{  1   2})_4
	- (e_{  2   1})_2 (e_{  1   3})_3 (e_{  3   2})_4
	+ (e_{  1   1})_2 (e_{  2   3})_3 (e_{  3   2})_4 \\
	&   - (e_{  3   2})_2 (e_{  1   1})_3 (e_{  2   3})_4
	+ (e_{  1   2})_2 (e_{  3   1})_3 (e_{  2   3})_4
	+ (e_{  3   1})_2 (e_{  1   2})_3 (e_{  2   3})_4
	- (e_{  1   1})_2 (e_{  3   2})_3 (e_{  2   3})_4 \\
	&   + (e_{  2   3})_1 (e_{  3   2})_{2} (e_{  1   1})_3
	+ (e_{  3   3})_1 (e_{  1   2})_{2} (e_{  2   1})_3
	- (e_{  1   3})_1 (e_{  3   2})_{2} (e_{  2   1})_3
	- (e_{  2   3})_1 (e_{  1   2})_{2} (e_{  3   1})_3  \\
	&	- (e_{  3   2})_1 (e_{  2   3})_{2} (e_{  1   1})_3
	+ (e_{  1   2})_1 (e_{  2   3})_{2} (e_{  3   1})_3
	- (e_{  3   3})_1 (e_{  2   1})_{2} (e_{  1   2})_3
	+ (e_{  1   3})_1 (e_{  2   1})_{2} (e_{  3   2})_3 \\
	&	+ (e_{  3   1})_1 (e_{  2   3})_{2} (e_{  1   2})_3
	- (e_{  1   1})_1 (e_{  2   3})_{2} (e_{  3   2})_3
	+ (e_{  3   2})_1 (e_{  2   1})_{2} (e_{  1   3})_3
	- (e_{  1   2})_1 (e_{  2   1})_{2} (e_{  3   3})_3 \\
	& - (e_{  2   1})_1 (e_{  3   2})_{2} (e_{  1   3})_3
	- (e_{  3   1})_1 (e_{  1   2})_{2} (e_{  2   3})_3
	+ (e_{  1   1})_1 (e_{  3   2})_{2} (e_{  2   3})_3
	+ (e_{  2   1})_1 (e_{  1   2})_{2} (e_{  3   3})_3 ,  
	\end{align*}
\begin{align*}
V_{L-1}^D= &- (e_{  2   3})_{L-1} (e_{  2   1})_L
+ (e_{  3   2})_{L-1} (e_{  1   2})_L
+ (e_{  1   2})_{L-1} (e_{  3   2})_L
- (e_{  2   1})_{L-1} (e_{  2   3})_L   \\
&- (e_{  2   3})_{L-2} (e_{  2   1})_{L-1}
+ (e_{  3   2})_{L-2} (e_{  1   2})_{L-1}
+ (e_{  1   2})_{L-2} (e_{  3   2})_{L-1}
- (e_{  2   1})_{L-2} (e_{  2   3})_{L-1},  	
\end{align*}
\begin{align*}
V_{L-1}^T= &   + (e_{  2   3})_{L-2} (e_{  3   2})_{{L-1}} (e_{  1   1})_L
+ (e_{  3   3})_{L-2} (e_{  1   2})_{{L-1}} (e_{  2   1})_L
- (e_{  1   3})_{L-2} (e_{  3   2})_{{L-1}} (e_{  2   1})_L
- (e_{  2   3})_{L-2} (e_{  1   2})_{{L-1}} (e_{  3   1})_L  \\
&	- (e_{  3   2})_{L-2} (e_{  2   3})_{{L-1}} (e_{  1   1})_L
+ (e_{  1   2})_{L-2} (e_{  2   3})_{{L-1}} (e_{  3   1})_L
- (e_{  3   3})_{L-2} (e_{  2   1})_{{L-1}} (e_{  1   2})_L
+ (e_{  1   3})_{L-2} (e_{  2   1})_{{L-1}} (e_{  3   2})_L \\
&	+ (e_{  3   1})_{L-2} (e_{  2   3})_{{L-1}} (e_{  1   2})_L
- (e_{  1   1})_{L-2} (e_{  2   3})_{{L-1}} (e_{  3   2})_L
+ (e_{  3   2})_{L-2} (e_{  2   1})_{{L-1}} (e_{  1   3})_L
- (e_{  1   2})_{L-2} (e_{  2   1})_{{L-1}} (e_{  3   3})_L \\
&    - (e_{  2   1})_{L-2} (e_{  3   2})_{{L-1}} (e_{  1   3})_L
- (e_{  3   1})_{L-2} (e_{  1   2})_{{L-1}} (e_{  2   3})_L
+ (e_{  1   1})_{L-2} (e_{  3   2})_{{L-1}} (e_{  2   3})_L
+ (e_{  2   1})_{L-2} (e_{  1   2})_{{L-1}} (e_{  3   3})_L  \\
&		- (e_{  2   3})_{{L-1}-2} (e_{  3   2})_{L-2} (e_{  1   1})_{L-1}
+ (e_{  2   3})_{{L-1}-2} (e_{  1   2})_{L-2} (e_{  3   1})_{L-1}
+ (e_{  3   2})_{{L-1}-2} (e_{  2   3})_{L-2} (e_{  1   1})_{L-1}
- (e_{  3   2})_{{L-1}-2} (e_{  1   3})_{L-2} (e_{  2   1})_{L-1}  \\
&   + (e_{  1   2})_{{L-1}-2} (e_{  3   3})_{L-2} (e_{  2   1})_{L-1}
- (e_{  1   2})_{{L-1}-2} (e_{  2   3})_{L-2} (e_{  3   1})_{L-1}
+ (e_{  2   3})_{{L-1}-2} (e_{  3   1})_{L-2} (e_{  1   2})_{L-1}
- (e_{  2   3})_{{L-1}-2} (e_{  1   1})_{L-2} (e_{  3   2})_{L-1}  \\
&   - (e_{  2   1})_{{L-1}-2} (e_{  3   3})_{L-2} (e_{  1   2})_{L-1}
+ (e_{  2   1})_{{L-1}-2} (e_{  1   3})_{L-2} (e_{  3   2})_{L-1}
- (e_{  3   2})_{{L-1}-2} (e_{  2   1})_{L-2} (e_{  1   3})_{L-1}
+ (e_{  3   2})_{{L-1}-2} (e_{  1   1})_{L-2} (e_{  2   3})_{L-1} \\
& 	- (e_{  1   2})_{{L-1}-2} (e_{  3   1})_{L-2} (e_{  2   3})_{L-1}
+ (e_{  1   2})_{{L-1}-2} (e_{  2   1})_{L-2} (e_{  3   3})_{L-1}
+ (e_{  2   1})_{{L-1}-2} (e_{  3   2})_{L-2} (e_{  1   3})_{L-1}
- (e_{  2   1})_{{L-1}-2} (e_{  1   2})_{L-2} (e_{  3   3})_{L-1},
\end{align*}
\begin{align*}
	V_L^D=&- (e_{  2   3})_{L-1} (e_{  2   1})_L
	+ (e_{  3   2})_{L-1} (e_{  1   2})_L
	+ (e_{  1   2})_{L-1} (e_{  3   2})_L
	- (e_{  2   1})_{L-1} (e_{  2   3})_L,	
\end{align*}
\begin{align*}
 V_L^T=&- (e_{  2   3})_L-2 (e_{  3   2})_L-1 (e_{  1   1})_L
 + (e_{  2   3})_L-2 (e_{  1   2})_L-1 (e_{  3   1})_L
 + (e_{  3   2})_L-2 (e_{  2   3})_L-1 (e_{  1   1})_L
 - (e_{  3   2})_L-2 (e_{  1   3})_L-1 (e_{  2   1})_L  \\
 &   + (e_{  1   2})_L-2 (e_{  3   3})_L-1 (e_{  2   1})_L
 - (e_{  1   2})_L-2 (e_{  2   3})_L-1 (e_{  3   1})_L
 + (e_{  2   3})_L-2 (e_{  3   1})_L-1 (e_{  1   2})_L
 - (e_{  2   3})_L-2 (e_{  1   1})_L-1 (e_{  3   2})_L  \\
 &   - (e_{  2   1})_L-2 (e_{  3   3})_L-1 (e_{  1   2})_L
 + (e_{  2   1})_L-2 (e_{  1   3})_L-1 (e_{  3   2})_L
 - (e_{  3   2})_L-2 (e_{  2   1})_L-1 (e_{  1   3})_L
 + (e_{  3   2})_L-2 (e_{  1   1})_L-1 (e_{  2   3})_L \\
 & 	- (e_{  1   2})_L-2 (e_{  3   1})_L-1 (e_{  2   3})_L
 + (e_{  1   2})_L-2 (e_{  2   1})_L-1 (e_{  3   3})_L
 + (e_{  2   1})_L-2 (e_{  3   2})_L-1 (e_{  1   3})_L
 - (e_{  2   1})_L-2 (e_{  1   2})_L-1 (e_{  3   3})_L.
\end{align*}
As follows from the lemma, we are led to a set of degenerate ground states  $g_j|\psi_0\rangle$,
which take the form $\sum _{i=1}^L (1-2\delta_{i\;j}) |+_1\ldots +_{i-1} 0_i +_{i+1} \ldots +_L \rangle$, up to a multiplicative constant.

The explicit expressions for the commutators between the Hamiltonian $H$ and the local unitary operations $g_j$ for other degenerate ground states are available, if requested.

\end{widetext}


\begin{thebibliography}{10}
	
	\bibitem{goldstone} J. Goldstone, Nuovo Cimento \textbf{19}, 154 (1961);
	J. Goldstone, A. Salam, and S. Weinberg, Phys. Rev. \textbf{127}, 965 (1962).
	
	\bibitem{Nambu1}
	Y. Nambu and G. Jona-Lasinio, Phys. Rev. \textbf{122}, 345 (1961).	
	
	
	
	\bibitem{Hnielsen} H. B. Nielsen and S. Chadha, Nucl. Phys. B \textbf{105}, 445 (1976).
	
	\bibitem{schafer}  T. Schafer, D. T. Son, M. A. Stephanov, D. Toublan, and J. J. M. Verbaarschot, Phys. Lett. B \textbf{522}, 67 (2001).
	
	\bibitem{miransky} V. A. Miransky and I. A. Shovkovy, Phys. Rev. Lett. \textbf{88}, 111601 (2002).
	
	\bibitem{nambu} Y. Nambu, J. Stat. Phys. \textbf{115}, 7 (2004).
	
	\bibitem{nicolis}  A. Nicolis and F. Piazza, Phys. Rev. Lett. \textbf{110}, 011602 (2013).
	
	\bibitem{brauner1}
	H. Watanabe, T. Brauner, and H. Murayama, Phys. Rev. Lett. \textbf{111}, 021601 (2013).
	
	
	\bibitem{brauner-watanabe} H. Watanabe and T. Brauner, Phys. Rev. D \textbf{84}, 125013 (2011).
	
	\bibitem{watanabe} H. Watanabe and H. Murayama, Phys. Rev. Lett. \textbf{108}, 251602 (2012); H. Watanabe and H. Murayama, Phys. Rev. X \textbf{4}, 031057 (2014).
	
	\bibitem{NG} Y. Hidaka, Phys. Rev. Lett. \textbf{110}, 091601 (2013);
	T. Hayata and Y. Hidaka, Phys. Rev. D \textbf{91}, 056006 (2015).
	
	\bibitem{NG1} D. A. Takahashi and M. Nitta, Ann. Phys. \textbf{354}, 101 (2015).	
	
	\bibitem{anderson} P. W. Anderson, Phys. Today \textbf{43}, 5, 117 (1990).
	
	\bibitem{peierls}  R. Peierls, J. Phys. A: Math. Gen. \textbf{24}, 5273 (1991);
	R. Peierls,	T. A. Kaplan, and P. W. Anderson, Phys. Today \textbf{44}, 2, 13 (1991).

     \bibitem{FMGM}  Q.-Q. Shi, Y.-W. Dai, H.-Q. Zhou, and I. P. McCulloch, arXiv: 2201.01071 (2022).
     
     \bibitem{LLspin1} Q.-Q. Shi, Y.-W. Dai, S.-H. Li, and H.-Q. Zhou, arXiv: 2204.05692 (2022).
     
     \bibitem{golden} H.-Q. Zhou, Q.-Q. Shi, I. P. McCulloch, and M. T. Batchelor, arXiv: 2302.13126 (2023).
     
    \bibitem{SU4} Q.-Q. Shi, H.-Q. Zhou, I. P. McCulloch, and  M. T. Batchelor, arXiv: 2309.04973 (2023).	

	\bibitem{Kallin} A. B. Kallin, M. B. Hastings, R. G. Melko, and	R. R. P. Singh, Phys. Rev. B \textbf{84}, 165134 (2011).
   
 
	\bibitem{Song}
	H. F. Song, N. Laflorencie, S. Rachel, and K. Le Hur, Phys. Rev. B \textbf{83}, 224410 (2011).
	
	\bibitem{Metlitski} M. A. Metlitski and T. Grover, arXiv: 1112.5166 (2015).
	
	\bibitem{squarecubic} S. Humeniuk and T. Roscilde, Phys. Rev. B \textbf{86}, 235116 (2012);
	I. Fr\'{e}rot and T. Roscilde, Phys. Rev. B \textbf{92}, 115129 (2015).
	
	\bibitem{bilayer} J. Helmes and S. Wessel,  Phys. Rev. B \textbf{89}, 245120
	(2014).
	
	\bibitem{typeAGM}   B. Kulchytskyy, C. M. Herdman, S. Inglis, and R. G. Melko, Phys. Rev. B \textbf{92}, 115146 (2015).
	
	\bibitem{alet} D. J. Luitz, X. Plat, F. Alet, and N. Laflorencie, Phys. Rev B \textbf{91}, 155145 (2015);
	N. Laflorencie, D. J. Luitz, and F. Alet, 	Phys. Rev. B \textbf{92}, 115126 (2015).

	\bibitem{bauer} D.-V. Bauer and J. O. Fjaerestad, Phys. Rev. B \textbf{101}, 195124 (2020).

    \bibitem{tasaki}  H. Tasaki, Physics and Mathematics of Quantum Many-Body Systems (Springer, 2020).
	
     \bibitem{baxterbook} R. J. Baxter, Exactly Solved Models in Statistical Mechanics (Academic Press, London, 1982).
	
	\bibitem{sutherlandb} B. Sutherland, {Beautiful Models, 70 Years of Exactly Solved Quantum Many-Body Problems} (World Scientific, Singapore, 2004).
	
	\bibitem{mccoy} B. M. McCoy, {Advanced Statistical Physics} (Oxford University Press, Oxford, 2009).		
	
	\bibitem{tla} H. N. V. Temperley and E. Lieb, Proc. R. Soc. Lond. \textbf{A322}, 251 (1971).
	
	\bibitem{martin} P. Martin, Potts Models and Related Problems in Statistical Mechanics (World Scientific, 1991).	
	
	\bibitem{katsura}  H. Watanabe, H. Katsura, and J. Y. Lee, arXiv: 2310.16881 (2023); J. Wouters, H. Katsura, and D. Schuricht, Phys. Rev. B \textbf{98}, 155119 (2018).
	
	 \bibitem{svd} M. A. Nielsen and I. L. Chuang, Quantum Computation and Quantum Information (Cambridge University Press, Cambridge, UK, 2000).
	
	
	\bibitem{finitesize} H.-Q. Zhou, Q.-Q. Shi, I. P. McCulloch, and M. T. Batchelor, arXiv: 2304.11339 (2023).
	
	\bibitem{doyon} O. A. Castro-Alvaredo and B. Doyon, Phys. Rev. Lett. \textbf{108}, 120401 (2012).
	
	\bibitem{popkov} V. Popkov and M. Salerno, Phys. Rev. A \textbf{71}, 012301 (2005); V. Popkov, M. Salerno, G. Sch\"{u}tz, Phys. Rev. A \textbf{72}, 032327 (2005).
	
	\bibitem{dtmodel}
	Y.-T. Oh, H. Katsura, H.-Y. Lee, and J. H. Han, Phys. Rev. B, \textbf{96}, 165126 (2017); H. Katsura, JPSJ News Comments \textbf{18}, 04 (2021).
	
	\bibitem{dtmodel2} T. Mashiko, S. Moriya, and K. Nomura, J. Phys. Soc. Jpn. \textbf{90}, 024005 (2021).
	
    \bibitem{AKLT}
	T. Affeck, T. Kennedy, E. Lieb, and H. Tasaki, Phys. Rev. Lett. \textbf{59}, 799 (1987).
	
	\bibitem{Sutherland}
	G. V. Uimin, JETP Lett. \textbf{12}, 225 (1970); C. K. Lai, J. Math. Phys. \textbf{15}, 1675 (1974); B. Sutherland, Phys.
	Rev. B \textbf{12}, 3795 (1975).
	
	\bibitem{TB} L. Takhtajan, Phys. Lett. A \textbf{87}, 479 (1982);
	H. Babujian, Nucl. Phys. B \textbf{215}, 317 (1983).
	
	\bibitem{barber} M. N. Barber and M. T. Batchelor, Phys. Rev. B \textbf{40} 4621 (1989).
	
	\bibitem{Chubukov} A. V. Chubukov, Phys. Rev. B \textbf{43}, 3337 (1991).
	
	\bibitem{Fath} G. F\'{a}th and J. S\'{o}lyom, Phys. Rev. B \textbf{44}, 11836 (1991);
	G. F\'{a}th and J. S\'{o}lyom, Phys. Rev. B \textbf{51}, 3620 (1995);
	K. Buchta, G. F\'{a}th, O. Legeza, and J. S\'{o}lyom, Phys. Rev. B \textbf{72}, 054433 (2005).
	
	\bibitem{Kawashima}
	N. Kawashima, Prog. Theor. Phys. Suppl. \textbf{145}, 138 (2002).
	
	\bibitem{Ivanov}
	B. A. Ivanov and A. K. Kolezhuk, Phys. Rev. B \textbf{68}, 052401 (2003).
	
	\bibitem{Rizzi} M. Rizzi, D. Rossini, G. De Chiara, S. Montangero, and R. Fazio, Phys. Rev. Lett. \textbf{95}, 240404 (2005).
	
	\bibitem{Lauchli} A. L\"{a}uchli, G. Schmid, and S. Trebst, Phys. Rev. B \textbf{74}, 144426 (2006).
	
	\bibitem{Porras} D. Porras, F. Verstraete, and J. I. Cirac, Phys. Rev. B \textbf{73}, 014410 (2006).
	
	
	\bibitem{Romero}
	O. Romero-Isart, K. Eckert, and A. Sanpera, Phys. Rev. A \textbf{75}, 050303 (2007).
	
	
	\bibitem{ronny} R. Thomale, S. Rachel, B. A. Bernevig, and D. P. Arovas, J. Stat. Mech. P07017 (2015); R. Lundgren,  {\it et al.}, Phys. Rev. B \textbf{94}, 081112 (2016).
	
	\bibitem{Rakov}
	M. V. Rakov and M. Weyrauch, J. Phys. Commun \textbf{1}, 015007 (2017).
	
	\bibitem{Sierra} J. I. Cirac and G. Sierra, Phys. Rev. B \textbf{81}, 104431 (2010);
	A. E. B. Nielsen, J. I. Cirac, and G. Sierra, J. Stat. Mech. P11014 (2011).
	
	
	\bibitem{daibb}
	Y.-W. Dai, Q.-Q. Shi, H.-Q. Zhou, and I. P. McCulloch, arXiv: 2201.01434 (2022).
	
	\bibitem{spins} B. Aufgebauer and A. Kl\"{u}mper,  J. Stat. Mech. P05018 (2010).
	
	\bibitem{saleur} N. Read and H. Saleur, Nucl. Phys. B, \textbf{777}, 263-315 (2007).
	
	\bibitem{moudgalya} S. Moudgalya and O. I. Motrunich, Phys. Rev. X  \textbf{12}, 011050 (2022).
	
	\bibitem{exactmps}	H.-Q. Zhou {\it et al.}, Exact matrix product state representations for a type of scale invariant states.

     \bibitem{binet}  R. Honsberger, Mathematical Gems, Vol 1, (MAA Press, Washington, D. C., 1974) p171.

\end{thebibliography}
\end{document}